\documentclass[preprint]{elsarticle}

\usepackage{lineno,hyperref}
\modulolinenumbers[5]
\usepackage{graphicx}
\usepackage{float}
\usepackage{setspace}
\usepackage{amsmath}
\usepackage{hyperref}
\usepackage{tabularx}
\usepackage{xcolor}

\journal{Nuclear Instruments and Methods}

\begin{document}

\begin{frontmatter}
\title{Multi-site Event Discrimination in Large Liquid Scintillation Detectors}  


\makeatletter
\def\@author#1{\g@addto@macro\elsauthors{\normalsize%
    \def\baselinestretch{1}%
    \upshape\authorsep#1\unskip\textsuperscript{%
      \ifx\@fnmark\@empty\else\unskip\sep\@fnmark\let\sep=,\fi
      \ifx\@corref\@empty\else\unskip\sep\@corref\let\sep=,\fi
      }%
    \def\authorsep{\unskip,\space}%
    \global\let\@fnmark\@empty
    \global\let\@corref\@empty  
    \global\let\sep\@empty}%
    \@eadauthor={#1}
}
\makeatother

\author{Jack Dunger\corref{mycorrespondingauthor}}\cortext[mycorrespondingauthor]{Corresponding author}
\ead{jackdunger@gmail.com}
\author{Steven D. Biller}
\address{Department of Physics, Univeristy of Oxford, Oxford OX1 3RH, UK}




\begin{abstract}

Simulation studies have been carried out to explore the ability to discriminate between single-site and multi-site energy depositions in large scale liquid scintillation detectors.  A robust approach has been found that is predicted to lead to a significant statistical separation for a large variety of event classes, providing a powerful tool to discriminate against backgrounds and break important degeneracies in signal extraction. This has particularly relevant implications for liquid scintillator searches for neutrinoless double beta decay ($0\nu\beta\beta$) from $^{130}$Te and $^{136}$Xe, where it is possible for a true $0\nu\beta\beta$ signal to be distinguished from most radioactive backgrounds (including those from cosmogenic production) as well as unknown gamma lines from the target isotope. 

\end{abstract}

\begin{keyword}
Liquid scintillator \sep particle ID \sep neutrinoless double beta decay
\end{keyword}

\end{frontmatter}

\section{Introduction}

Many of the physics signals targeted by large scale liquid scintillation detectors, such as solar neutrinos or neutrinoless double beta decay ($0\nu\beta\beta$), only involve charged particles that deposit their energy over a relatively short length of track. By contrast, many of the relevant backgrounds often involve the production of gamma rays that can travel measurable distances before Compton scattering, thus producing ionisation at separated sites. The separation between such single-site and multi-site event classes has been utilised to great effect in a number of other detector technologies \cite{exo_ms,Agostini2013} but has so far not been particularly strongly exploited in liquid scintillator, except for ortho-positronium events \cite{Bellini:2013lnn} and some limited exploration of other specific backgrounds \cite{Dunger_2017,Li:2018rzw}. Modern large scale liquid scintillation detectors with large photocathode coverage and relatively fast photomultiplier tubes (PMTs) can generally resolve the position of light generation from timing information to a precision smaller in scale than a typical Compton interaction length. Thus, the timing information carries significant potential to discriminate between single-site and multi-site events. Even the statistical separation of these classes can have a substantial impact in breaking important degeneracies in signal extraction.

The simulation studies presented here are based on a detector model very similar to SNO+, but the results are applicable to any existing or proposed large scintillator detector, including Borexino, KamLAND-Zen, JUNO, THEIA and the Jinping Neutrino Experiment. To aid such comparisons, we also consider variants of the model with different levels of light collection and PMTs with different transit time spreads (TTS).


\section{Model}

The analysis techniques presented here were tested on events simulated in a SNO+-like detector\footnote{The detector geometry closely matches the SNO+ detector but the number of functioning PMTs and the exact scintillator model will differ.}. A comprehensive description of the detector is given in \cite{Andringa:2015tza,Boger:1999bb}. This section outlines only the features most relevant to the study.

The target volume of the detector is a 12m diameter acrylic sphere housing $\approx$800t of liquid scintillator. Charged particles produced inside the target excite $\pi$ orbital electrons in nearby scintillator molecules. The relaxation of these electrons produces light along the particle track which is isotropic and emitted according to a time profile intrinsic to the scintillator. Here, the scintillator considered is Linear Alkyl Benzene (LAB) doped with 2g/L 2,5-Diphenyloxazole (PPO) and 15mg/L of 1,4-Bis(2-methylstyryl)benzene (bisMSB). The $\beta$ timing profile can be approximated by the sum of three exponentials:
\begin{equation}
	P(t_{em}) = 0.7\times e^{\frac{-(t_{em} - t_0)}{3.4\textrm{ns}}}
	+ 0.24\times e^{\frac{-(t_{em} - t_0)}{12.0\textrm{ns}}}
	+ 0.06\times e^{\frac{-(t_{em} - t_0)}{120.0\textrm{ns}}}
	\label{lab:eq_scint_resp}
\end{equation}
where $t_0$ is the excitation time and $t_{em}$ is the photon emission time.

Light produced in the scintillator is detected by 9300 inward facing 8" Hamamatsu R1408 PMTs installed 8.39m from the target centre. The PMTs have a peak efficiency\footnote{collection efficiency $\times$ quantum efficiency} of 13.5\% at 440nm and a transit time spread of 3.7ns FWHM. Reflective concentrators fitted to each PMT produce an effective photocathode-coverage of around 50\% \cite{Boger:1999bb}. For 2.5MeV events in the central 3.5m of the detector, the light yield is 380 PMT hits/MeV.

The R1408 is far from the cutting edge, so this work also considers what could be achieved if they were were replaced with more modern 8" variants (section~\ref{sec:upgraded_pmts}). The PMT referred to as HQE (High Quantum Efficiency) uses efficiency curves and charge distributions measured for the Hamamatsu R5912\cite{TheDEAP:2017bxf}. The HQE + FastTTS model further modifies the transit time distribution so that it has a prompt-peak TTS (transit time spread) of 1ns FWHM\footnote{the pre-pulsing and after-pulsing behaviour is the same as the R1408.}; prototypes with close to this performance have already been produced by Hamamatsu \cite{Kaptanoglu:2017jxo}.

Events were simulated using the SNO+ version of the \texttt{RAT} software package\cite{rat}. It contains a \texttt{GLG4Sim}\cite{glg4sim} simulation that handles the production of scintillation photons and a full \texttt{Geant4}\cite{geant4} detector simulation that individually tracks each photon through the geometry, accounting for reflection, refraction, scattering, absorption and re-emission. The PMTs and concentrators are modelled as 3D objects and the front-end and trigger system are simulated in full, including electronics noise. Radioactive decays are generated using a version of \texttt{Decay0}\cite{Ponkratenko:2000um} and custom code is used to handle ortho-positronium formation at the end of positron tracks \cite{jackthesis}.

\section{Event Classification}
\label{sec:classification}
Particles travelling through liquid scintillator form tracks whose length and shape are determined by the relevant energy loss processes. In any given event, there may be several individual tracks and the scintillation light produced along each will create an energy deposition region in time and space. If this region is much smaller than the detector's vertex resolution (typically 5-10cm, 1ns), the event will be indistinguishable from a point-like deposition and the event is considered `single-site'.  Conversely, if the size of the region is comparable to, or greater than, the vertex resolution, the detector will be sensitive to the exact form of the deposition and the event is considered `multi-site'. The following section categorises several important event sources as `single-site' or `multi-site'.

\subsection{Single-site}
\label{sec:single_site}

\begin{figure}
	\centering
	\includegraphics[width=.8\textwidth]{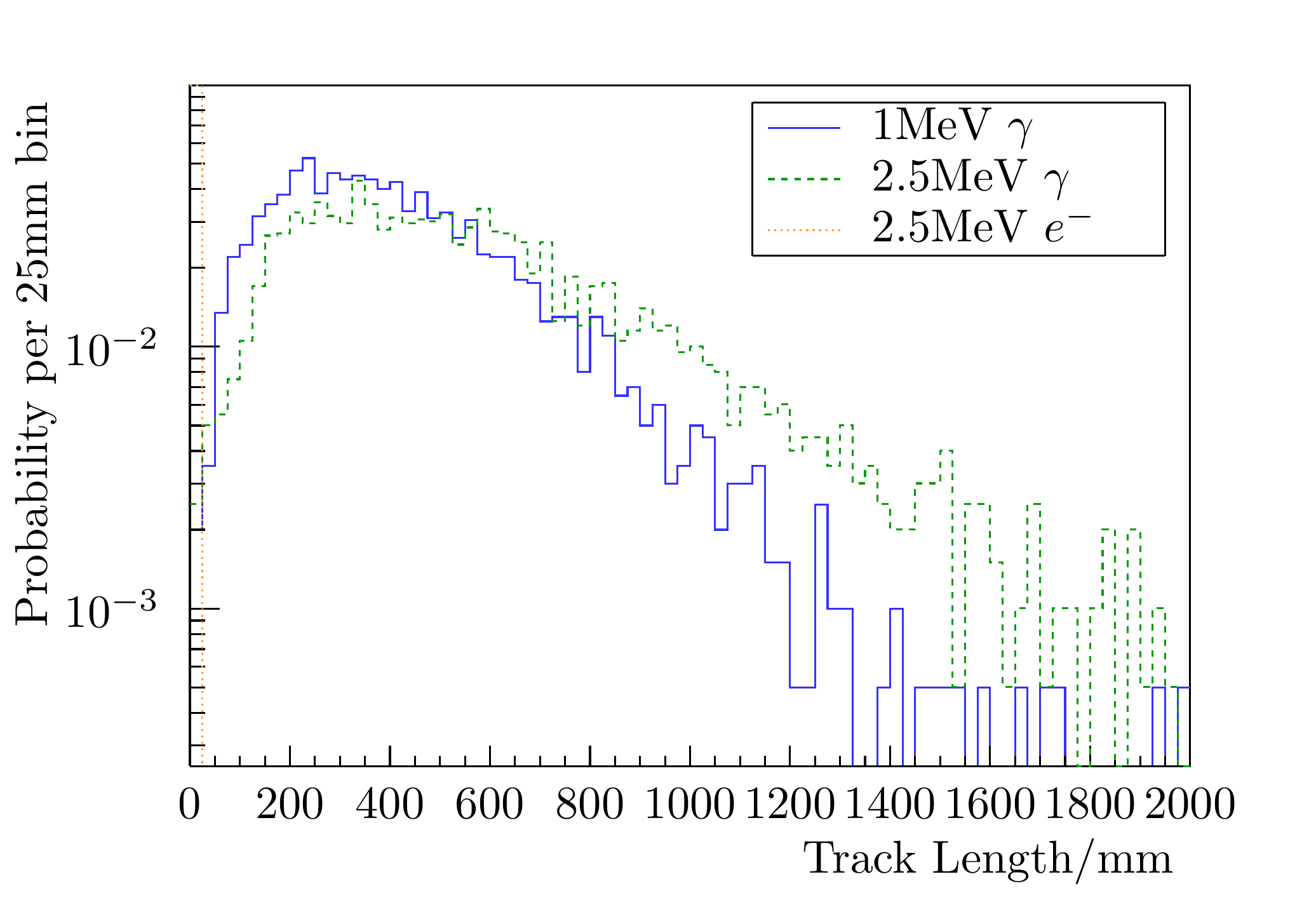}
	\caption{Simulated electron and gamma track lengths in liquid scintillator. The track length is defined as the distance between the start and the end of the track. \cite{jackthesis}}
	\label{fig:track_lens}
\end{figure}

Electron multiple scattering ensures that electrons at 2.5MeV, the $0\nu\beta\beta$ energy in $^{130}$Te and $^{136}$Xe, create tracks in liquid scintillator that are less than 1.5cm in length (figure~\ref{fig:track_lens}). This is significantly smaller than a typical vertex resolution, so these events appear  single-site. In liquid scintillator, solar neutrinos are primarily detected via the single electrons produced in $\nu-e$ elastic scattering, so the solar neutrino signal will be single-site. $2\nu\beta\beta$ and $0\nu\beta\beta$ events will also be single-site because vertex resolution makes two short electron tracks indistinguishable from one.

\subsection{Multi-site}

Chargeless $\gamma$ do not scintillate directly, but are detectable via secondaries produced in interactions with the scintillator. For a mostly carbon target, the interaction cross-section of $\mathcal{O}${}(1MeV) $\gamma$ is dominated by Compton scattering \cite{Agashe:2014kda}, which produces scintillating electrons.

In LAB, the Compton length for a 1MeV $\gamma$ is close to 18cm \cite{jackthesis}. This is notably larger than the typical vertex resolution so, even if $\gamma$s only scattered once, one would expect a measurable effect. In fact, each $\gamma$ scatters several times, extending the track past the first scatter. Indeed, figure~\ref{fig:track_lens} shows the track lengths of 1MeV and 2.5MeV $\gamma$; the averages are 43cm and 57cm, respectively. These are significantly larger than the vertex resolution and there is a tail that extends past 1m (20$\sigma_{vtx}$). The time between scatters is also significant: LAB has a refractive index of $\approx$ 1.5 for optical wavelengths, so 40cm corresponds to a time delay of $\sim$2ns. This is comparable to TTS modern PMTS (including each of those considered here) and the dominant scintillation decay constants.

A positron scintillates identically to an electron until the end of its track, where it will annihilate with a nearby electron to produce two 511keV $\gamma$. The multiple Compton scattering of these $\gamma$ will give these events a multi-site component. Furthermore, in some LAB cocktails loaded with tellurium, for example, there is a 36\% chance of producing a long-lived ortho-positronium (o-Ps) state that has a lifetime of 2.7ns \cite{PhysRevC.88.065502}. If o-Ps is formed at the end of the positron track, the annihilation $\gamma$ will be delayed by the decay time of o-Ps. The average delay is large compared with the vertex time resolution, and will be comfortably detectable in many cases.

Radioactive $\beta^\pm$ decays are often followed by one or more nuclear de-excitation $\gamma$ after picosecond time scales. For these events, the multi-site nature is actually more pronounced than pure $\gamma$. This is because the time taken to reach the first Compton scatter is typically the longest (the Compton length falls with energy) and so it produces the largest time delay. In pure $\gamma$ events, this largest delay is unobservable because light is not emitted until the first scatter occurs and so the event appears to start then. However, in $\beta^\pm\gamma$ decays, the $\beta$ emits light immediately and so the delay before the first scatter \textit{can} be observed, as the delay between the first light from the $\beta$ and the first light from the $\gamma$.

\section{Time Residuals and Discrimination Statistic}
\label{sec:tres_disc}

This section outlines how to practically exploit the timing differences described in section~\ref{sec:classification}. First, it shows that multi-site events have broader PDFs in the PMT hit times after they are corrected for photon time of flight. Second, it demonstrates that likelihood-ratio tests using these distributions can be used to statistically separate $e^-$ from $\gamma, e^+$ and $0\nu\beta\beta$ from more complicated $\beta^\pm\gamma$ decays.

\subsection{Time Residuals}
\label{sec:time_residual}

The PMT hit times are strongly dependent on the time and position of the event, but it is possible to remove this dependence, to first order, by correcting for the apparent time of flight of each photon. These `time residuals' are defined by equation~\ref{eq:tres_def}

\begin{equation}
	t_{res}^i \equiv t_{hit}^i - t_0 - t_{flight}^i
	\label{eq:tres_def}
\end{equation}
where $t_0$ is the global event time, relative to the trigger, and $t_{res}^i$, $t_{hit}^i$ and $t_{flight}^i$ are the time residual, hit time and calculated time of flight for the $i$th hit, respectively. In order to calculate time of flight, the event time and position must first be estimated. This is achieved using a maximum likelihood fit based on the simulated time profiles of 3MeV e$^-$ \cite{ianthesis,jackthesis}. Typical vertex fitters, including the one employed here, are tuned to reconstruct point-like depositions from short-range tracks, but they also work well for reconstructing an energy weighted average position for multi-site events such as $\gamma$ and $e^+$. On average, 2.5MeV electrons events are reconstructed 92mm from the true event position; for positrons with the same number of hits the distance is 98mm.

Up to inaccuracies caused by the PMT TTS, non-straight photon paths produced by reflection and refraction and finite vertex resolution, an event's time residuals are an estimate of the emission times of its scintillation photons. Therefore they should be sensitive to the single-site or multi-site nature of events.

\begin{figure}
	\includegraphics[width=\textwidth]{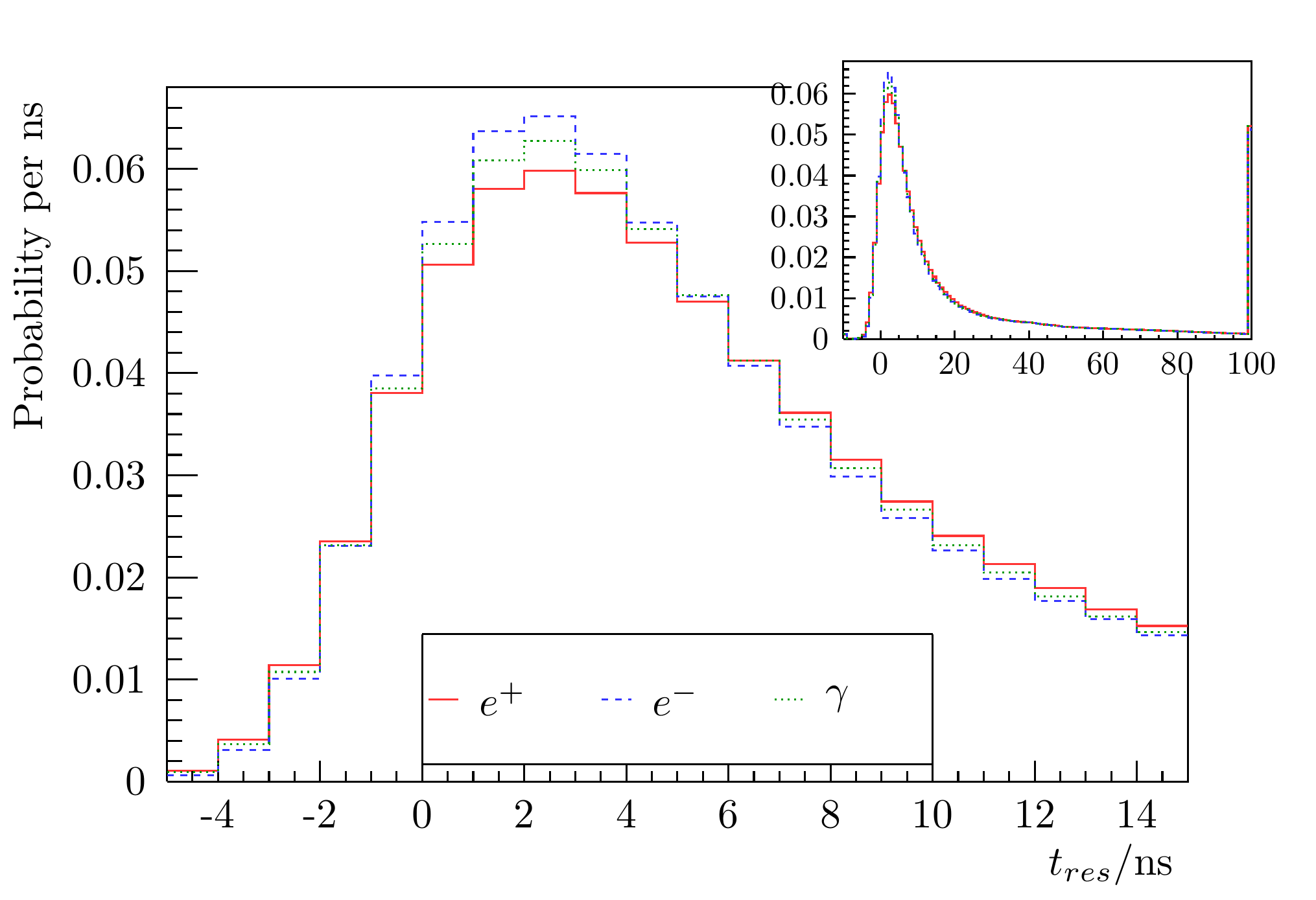}
	\caption{Simulated time residual distributions for $e^-$, $e^+$ and $\gamma$ uniformly generated inside the scintillator. $r_{fit} < 3.5\textrm{m}$, $2.31\textrm{MeV} < E_{fit} < 2.68\textrm{MeV}$. Particle energies were selected such that the three event classes had the same average number of PMT hits as a 2.5MeV $e^-$. Inset: the same distributions viewed over a wider range. The distributions have been averaged over 40,000 events and normalised to 1 hit.}
	\label{fig:tres_diff1}
\end{figure}

Figure~\ref{fig:tres_diff1} shows the time residuals of e$^-$, e$^+$ and $\gamma$ events in the scintillator. The $e^-$ distribution is close to the scintillator's emission curve because the deposition is point-like and instantaneous, but the distributions of $e^+$ and $\gamma$ events are broader. This happens for two reasons: first, the individual depositions are separated in time by $\gamma$ free-streaming and the o-Ps lifetime. Second, the depositions are separated in space around the reconstructed vertex. This broadens the PMT hit times in a way that cannot be accounted for by the time of flight correction in equation~\ref{eq:tres_def}, because it uses the reconstructed vertex as the sole point of origin for all of the photons.

It is important to recognise that, although differences in the individual distributions may appear to be modest, the sampling of these distributions in an event is large because modern detectors collect many hundreds of PMT hits per MeV, or more. The profile shape is therefore extremely well determined over the whole event, allowing significant discrimination power.

Note that the broadening is apparent on both the late \textit{and} early sides of the peak, even though o-Ps formation and $\gamma$ free-streaming can only make photon emission later. This is a reconstruction effect: by demanding that events match the expected $e^-$ hit times, the reconstruction algorithm always chooses a vertex that produces a peak at 3ns, where it appears for $e^-$ events. To achieve this, the time residuals of $\gamma$ and $e^+$ events are effectively shifted to earlier times, which produces an excess at small $t_{res}$.

\subsection{Discrimination}

The obvious way to test events against the distributions shown in figure~\ref{fig:tres_diff1} is to employ a likelihood-ratio, which guarantees optimal cut performance and is easy to calculate, provided the hits are independent\footnote{we found that accounting for correlations provided little benefit for these event types.}:

\begin{equation}
 	\Delta\log\mathcal{L} = \sum_{i = 0}^{N_{hit}}\log\frac{P_{S}(t_{res}^i)}{P_{B}(t_{res}^i)}
\end{equation}
where $t_{res}^i$ is the time residual of the $i$th hit, $N_{hit}$ is the number of hits and $P_{S/B}$ is the probability of observing a hit with time residual $t_{res}^i$ in a signal/background events, respectively (figure~\ref{fig:tres_diff1}). 

\begin{figure}
	\resizebox{\textwidth}{!}{
		\begin{tabular}{c}
			\includegraphics[width=.35\textwidth]{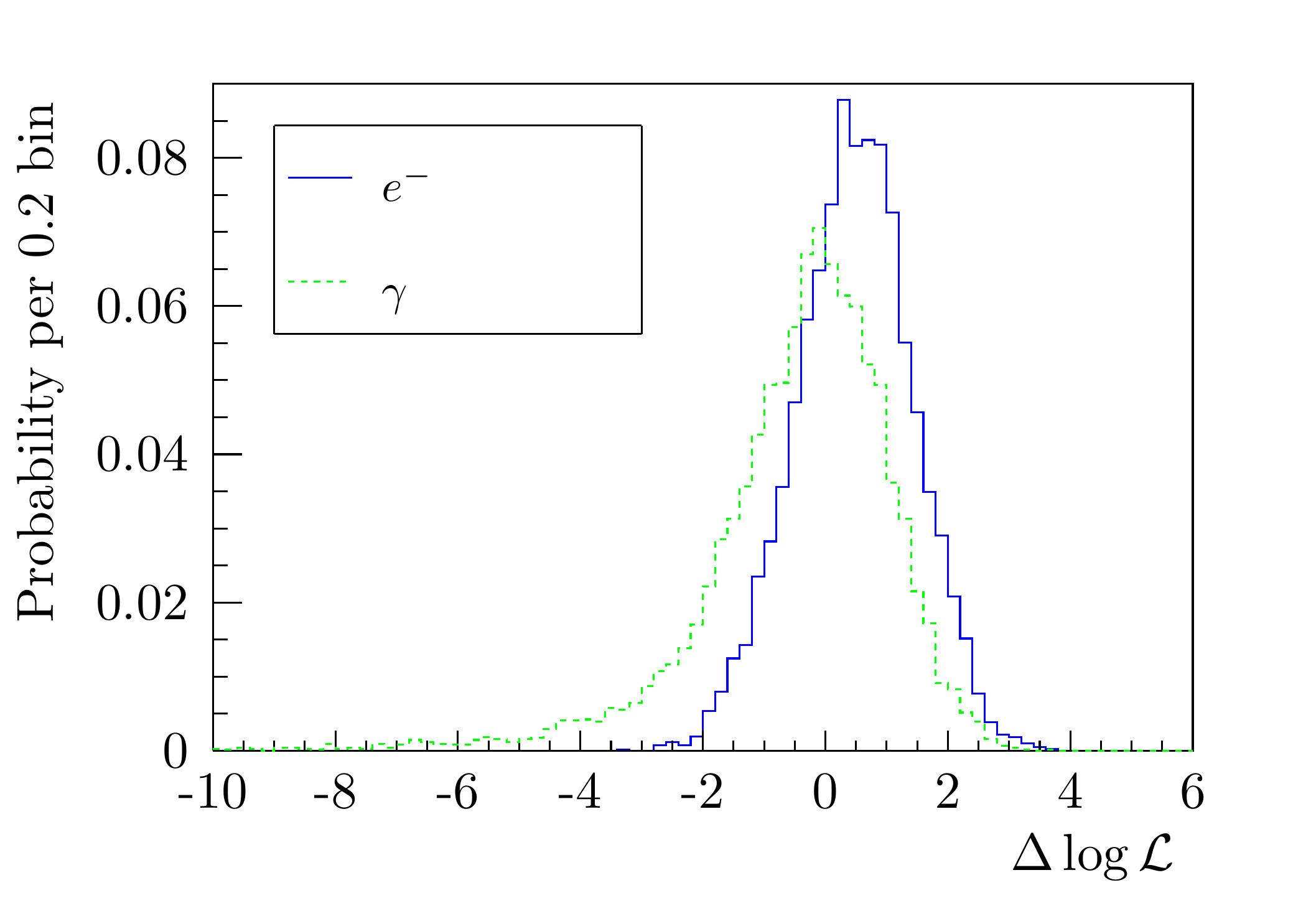} \\
			\includegraphics[width=.35\textwidth]{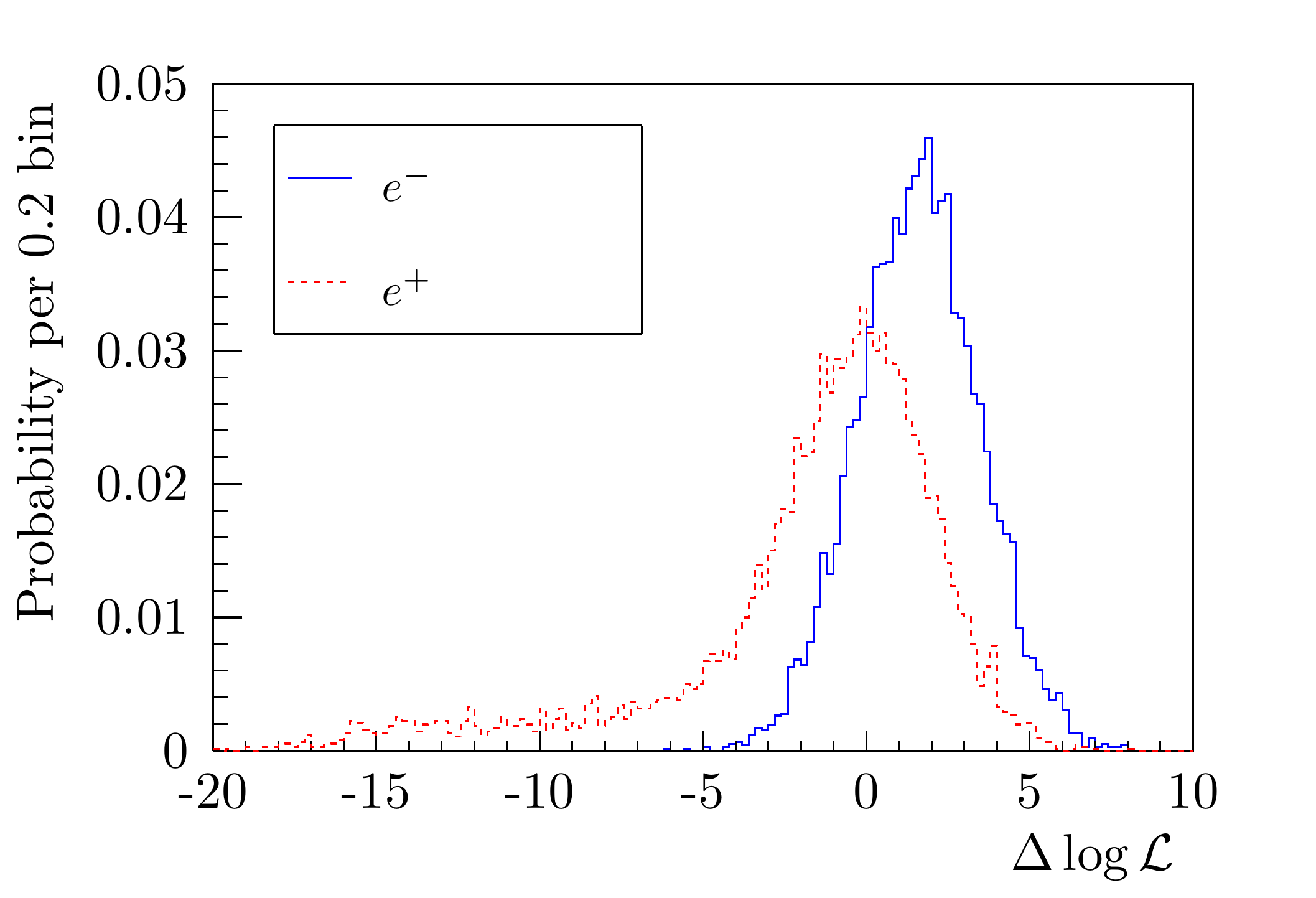} \\
		\end{tabular}
	}
	\caption{$\Delta\log\mathcal{L}$ distributions for discriminating $\gamma$ and $e^+$ from $e^-$. $r < 3500$mm, 2.31MeV $< E_{fit} < $ 2.68MeV. A PDF specific to the background considered was used in each case. Particle energies were selected such that the three event classes had the same average number of PMT hits as a 2.5MeV $e^-$.}
	\label{fig:delta_log_lh}
\end{figure}

Figure~\ref{fig:delta_log_lh} shows these distributions with  $S=e^-$, $B=e^+$/$\gamma$. The distributions do not permit efficient event-by-event separation, but the differences between the two populations can be well discriminated statistically. In both plots the populations are mostly Gaussian, but there is also a long tail on the $\gamma$ and $e^+$ distributions that are caused by non-Poissonian event-to-event variation. The events in the tails are those that have particularly long $\gamma$ tracks or particularly long-lived o-Ps states.

One could make use of the separations in figure~\ref{fig:delta_log_lh} by cutting on $\Delta\log\mathcal{L}$ to eliminate particles of a given type. In principle, one such discriminant would be required for each possible background, but the following section shows that they generalise well between radioactive backgrounds with similar topologies.

\subsection{Radioactive decays}
\label{sec:radioactive_decays}
Most naturally occurring backgrounds in the region of a few MeV have a much richer structure than single $\gamma$ or $e^+$ emission and often involve a number of different decay branches.
This section investigates a more practical use case: distinguishing single-site $0\nu\beta\beta$ events from multi-site backgrounds produced by the decay of isotopes resulting from the cosmogenic activation of tellurium. The most dangerous of these isotopes are those that have $\mathcal{O}$(1yr) half-lives and reconstructed energy spectra that overlap strongly with $0\nu\beta\beta$ in $^{130}$Te. Using these criteria, $^{22}$Na, $^{60}$Co $^{88}$Y, $^{214}$Sb, $^{44}$Sc and $^{110}$Ag(m) were identified as the most problematic isotopes (detailed discussions are given in \cite{jackthesis,Lozza:2014haa}). Each decay involves multiple branches, but the isotopes can be separated into two distinct groups by the type of the highest energy emitted particle(s). First, the $\gamma$ emitters: $^{60}$Co, $^{110}$Ag(m), $^{124}$Sb and $^{88}$Y are $\beta^-\gamma$ decays that, for energies near 2.5MeV, deposit a majority of their energy in $\gamma$s. Second, the $\beta^+\gamma$ emitters: $^{22}$Na and $^{44}$Sc are $\beta^+$ decays that produce $\gamma$s at 1.2MeV and 1.3MeV, respectively. In these decays, there are comparable deposits from both $\beta^+$ and $\gamma$.

\begin{figure}
	\resizebox{\textwidth}{!}{
	\begin{tabular}{c}
		\includegraphics[width=\textwidth]{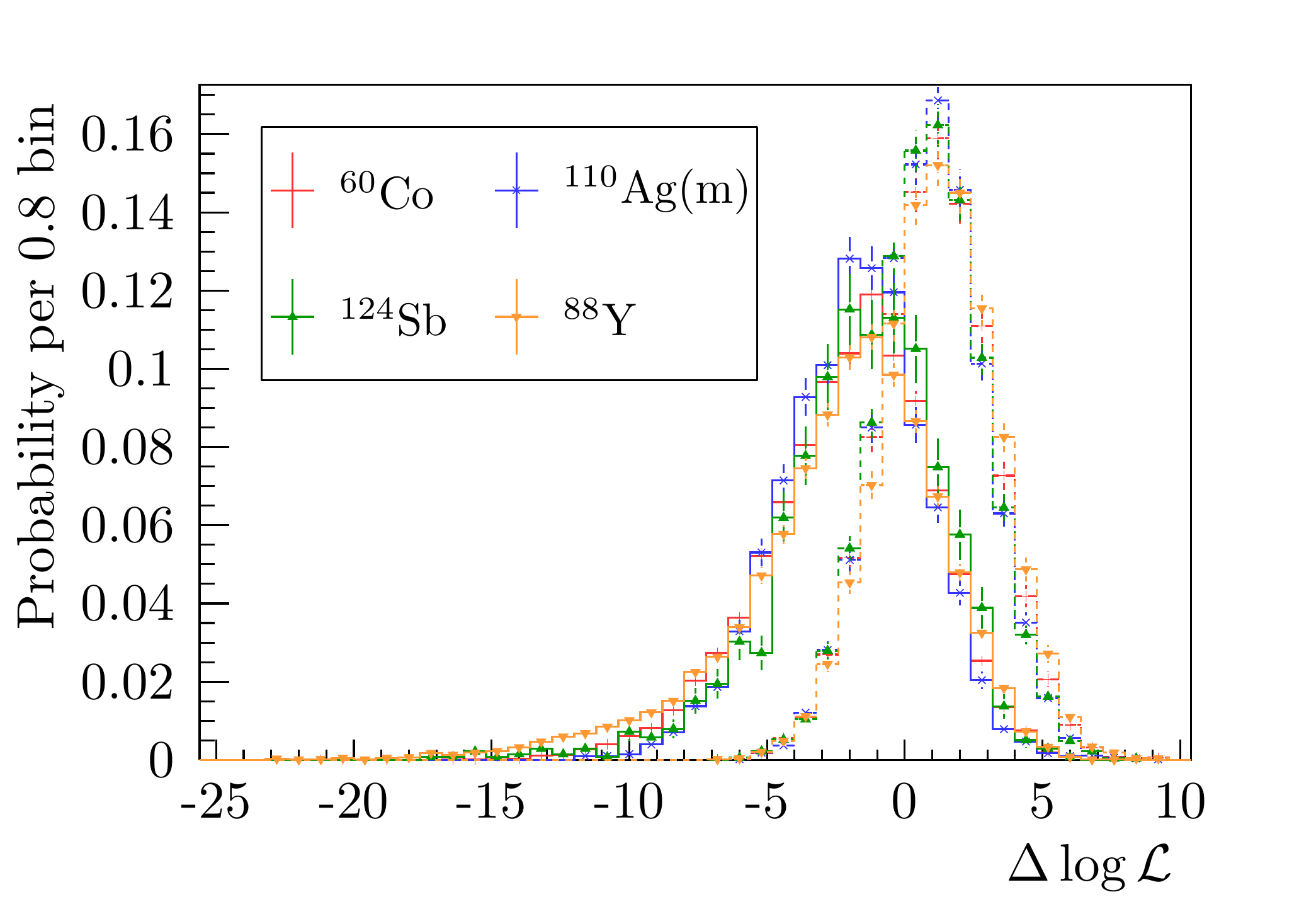} \\
		\includegraphics[width=\textwidth]{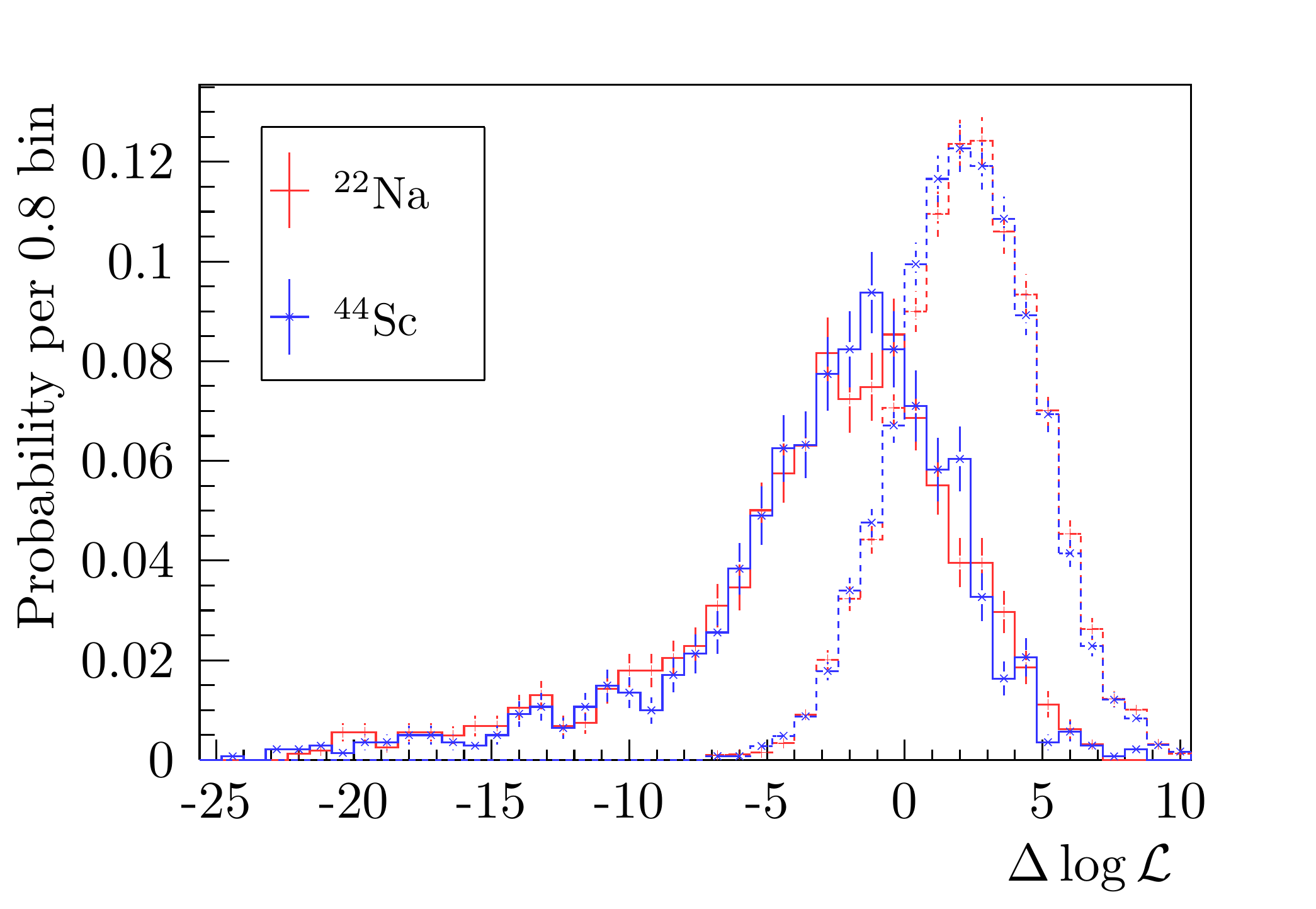} \\
	\end{tabular}
	}
	\caption{$\Delta\log\mathcal{L}$ distributions comparing cosmogenic isotope decays with $0\nu\beta\beta$. Dotted lines: $0\nu\beta\beta$ events. Solid lines: background events. The colour of the dotted $0\nu\beta\beta$ lines indicate the background event PDF used to calculate the likelihood ratio. Left: $\gamma$ dominated decays. Right: $\beta^+\gamma$ dominated decays. $r < 3500$mm, 2.31MeV $< E_{fit} < $ 2.68MeV.}
	\label{fig:cosmo_dlhs}
\end{figure}

Figure~\ref{fig:cosmo_dlhs} shows the $\Delta\log\mathcal{L}$ distributions for the backgrounds in these two groups, with $S=0\nu\beta\beta$. The $\gamma$ and $\beta^+\gamma$ dominated decays can be separated from $0\nu\beta\beta$ by 1 and 1.5 RMS widths, respectively. Most significantly, the variation \textit{between} the backgrounds in each group is very small. This demonstrates that the technique is robust: the separation power is dominated by the physics of the highest energy emitted particles rather than the exact details the decay. Indeed, using a single PDF for each group (testing $^{44}$Sc events using the $^{22}$Na PDF and $^{88}$Y events using a $^{60}$Co PDF etc.) leads to no observable cost in efficiency \cite{jackthesis} and even testing the $\beta^+\gamma$ events using a $\beta^-\gamma$ PDF still gives reasonable performance. Critically, this means that multi-site discrimination can be applied to a wide range of possible backgrounds without the need to tailor a discriminant for each.

\subsection{Dependence on event energy and radius}
\begin{figure}
	\resizebox{\textwidth}{!}{%
	\begin{tabular}{c c}
		\includegraphics[width=\textwidth]{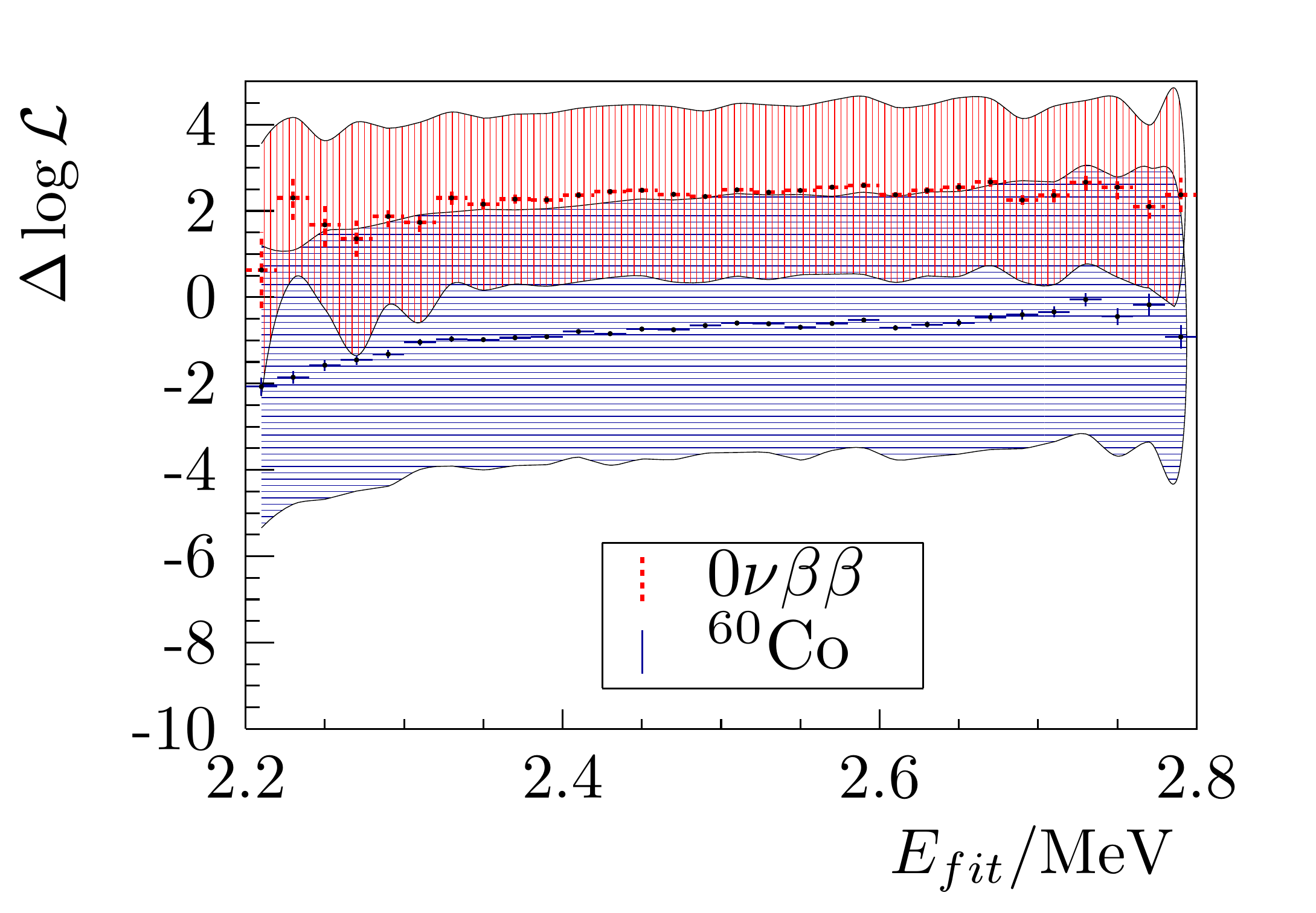} & 
		\includegraphics[width=\textwidth]{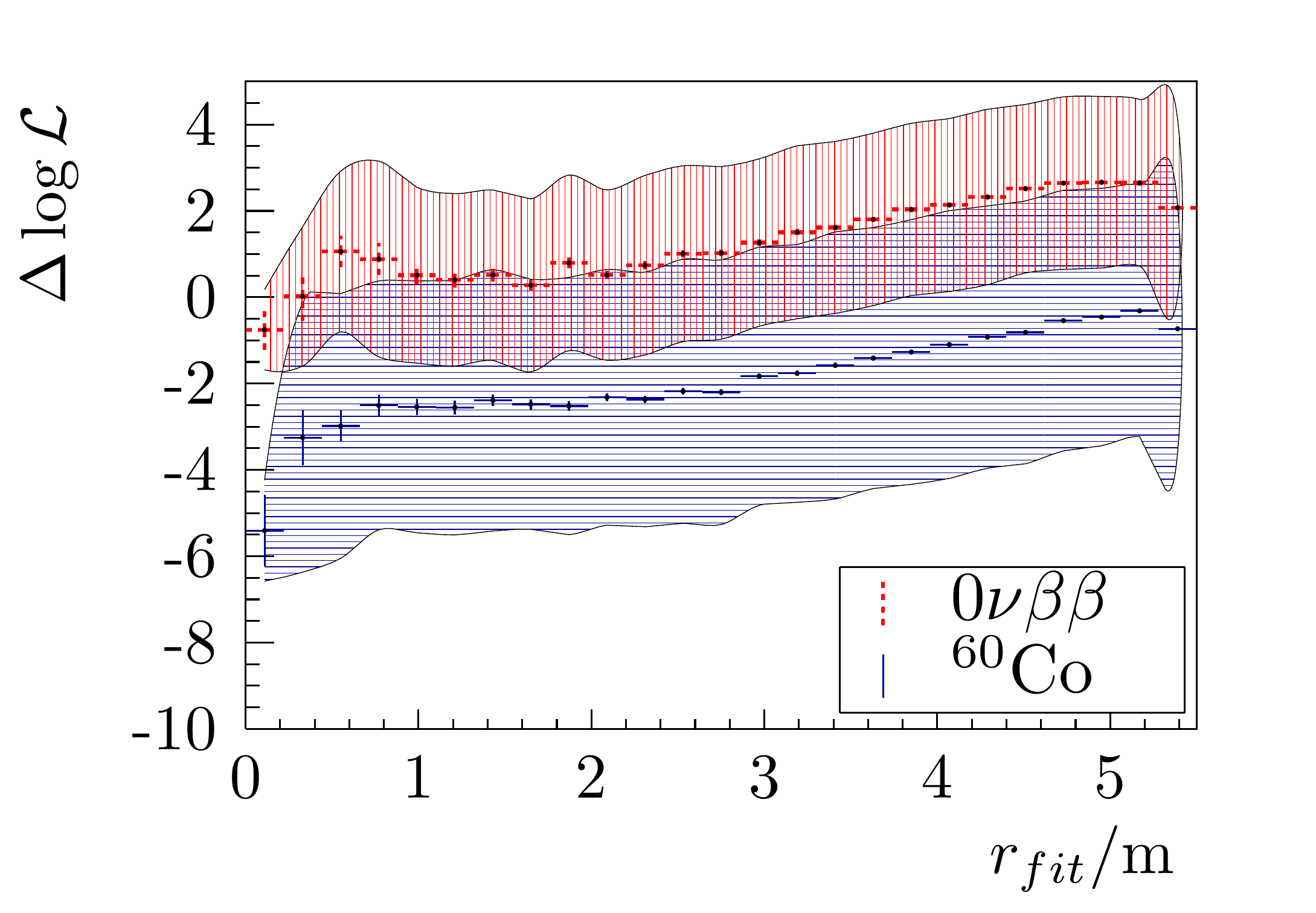} \\
		\includegraphics[width=\textwidth]{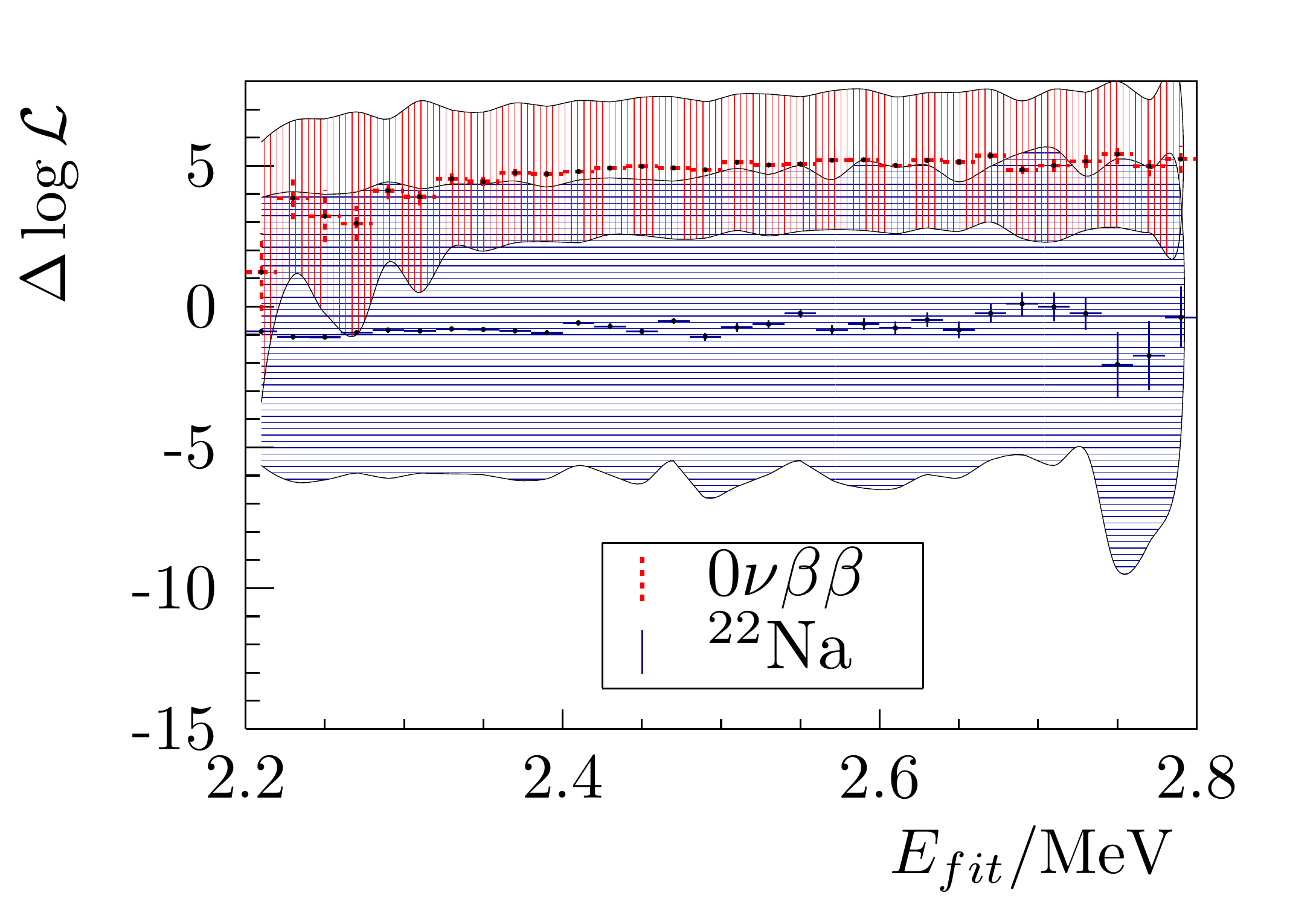} & 
		\includegraphics[width=\textwidth]{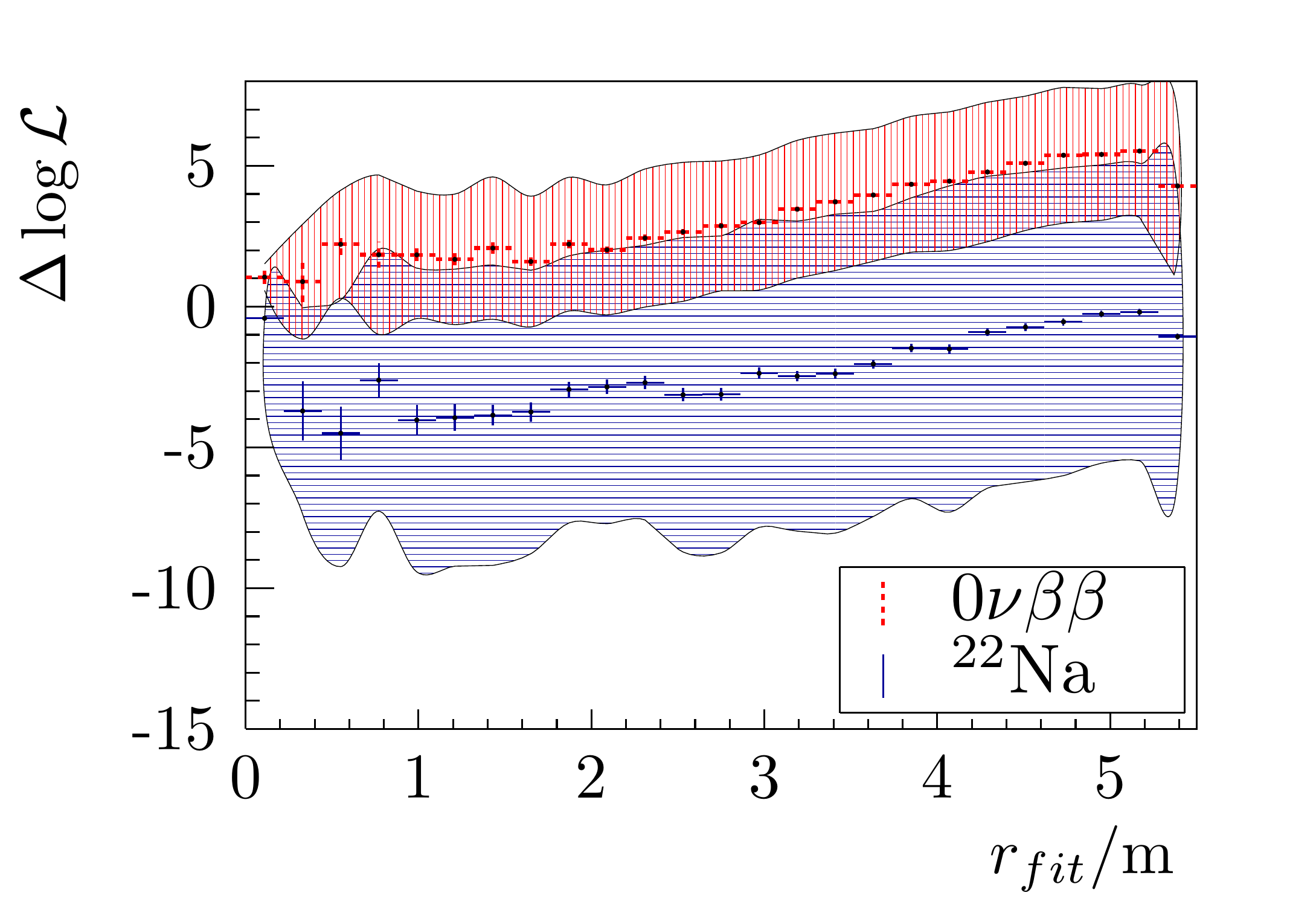} \\
	\end{tabular}
	}
	\caption{The dependence of $\Delta\log\mathcal{L}$ on reconstructed energy (left, $r_{fit} < 3.5$m) and reconstructed radius (right, $2.31\textrm{MeV} < E_{fit} < 2.68\textrm{MeV}$). Comparing $0\nu\beta\beta$ events with $^{60}$Co (top) and $^{22}$Na (bottom) events. The error bars are the standard error on the mean; the shaded region shows the RMS. The edge of the acrylic vessel sits at 6m.}
	\label{fig:llr_running}
\end{figure}

Exploiting $\Delta\log\mathcal{L}$ in signal extraction requires an understanding of how it depends on the event radius and energy, the typical observables used in a $0\nu\beta\beta$ likelihood fit. Figure~\ref{fig:llr_running} shows these dependencies explicitly for $S=0\nu\beta\beta$ and $B = ^{60}$Co, $^{22}$Na; the pattern is mirrored for other backgrounds. The patterns emerge because the shape of the time residuals for both $S$ and $B$ change with event energy and event radius, but the PDFs they are compared with in $\Delta\log\mathcal{L}$ are fixed averages over an energy window around 2.5MeV and $r < 3.5$m.

In particular, the time residuals for both $S$ and $B$ become more strongly peaked as the event position moves away from the detector centre. This is because solid angle and optical effects concentrate the PMT hits on the near side of the detector. These hits are from photons that have travelled shorter distances and are therefore less prone to optical effects that delay arrival at a PMT, relative to a straight line path. For this particular detector model, this trend is reversed above 5m: in this region there are total internal reflection effects that broaden the time residuals (more non-straight line paths) making both seem more background-like. 

There is a similar, but less pronounced, trend with event energy. More energetic events appear more signal-like because, at higher energies, a greater fraction of PMTs are hit by more than one photon. On average, these `multi-hits' produce earlier hit-times\footnote{The electronics modelled here do not resolve individual photo-electrons (p.e.) but rather measure the first time each PMT charge signal crosses a threshold. If several p.e. are created on a PMT, the threshold will be crossed, on average, earlier. To confirm this, it was explicitly verified that the creation times of individual p.e. do not depend on event energy.} and therefore more peaked time residuals. Critically, the separation of the $S$ and $B$ hypotheses is constant around the $0\nu\beta\beta$ region of interest in $^{130}$Te (2.5MeV).

\subsection{Upgraded PMTs}
\label{sec:upgraded_pmts}

It is useful to ask what could be achieved in future experiments with improved PMTs. Modern quantum efficiencies would allow better sampling of the distributions in figure~\ref{fig:tres_diff1} and faster PMTs would be able to better resolve the photon arrival times, increasing the difference between the distributions themselves.

\begin{figure}
	\resizebox{\textwidth}{!}{%
		\begin{tabular}{c c}
			\includegraphics[width=0.5\textwidth]{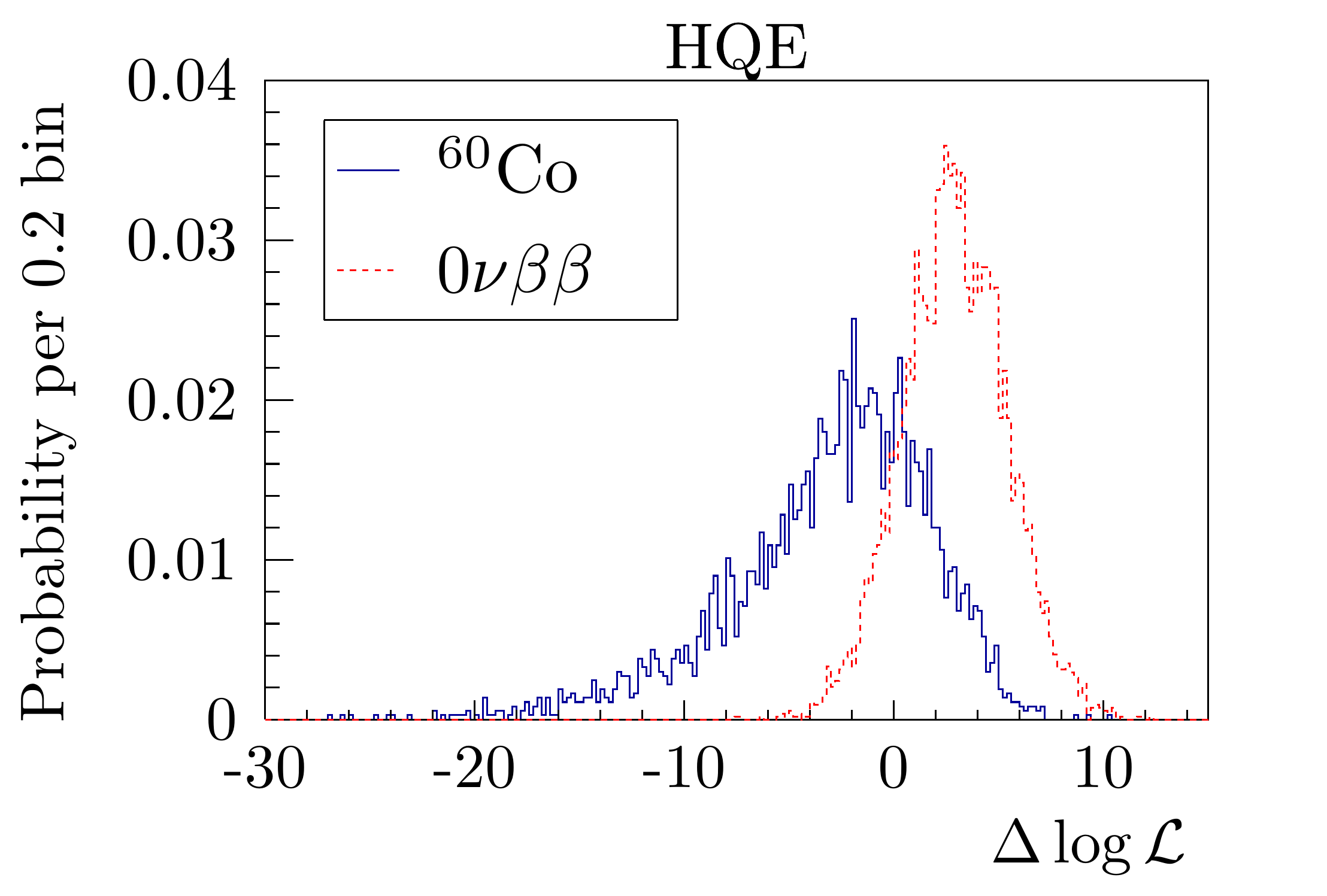} & 
			\includegraphics[width=0.5\textwidth]{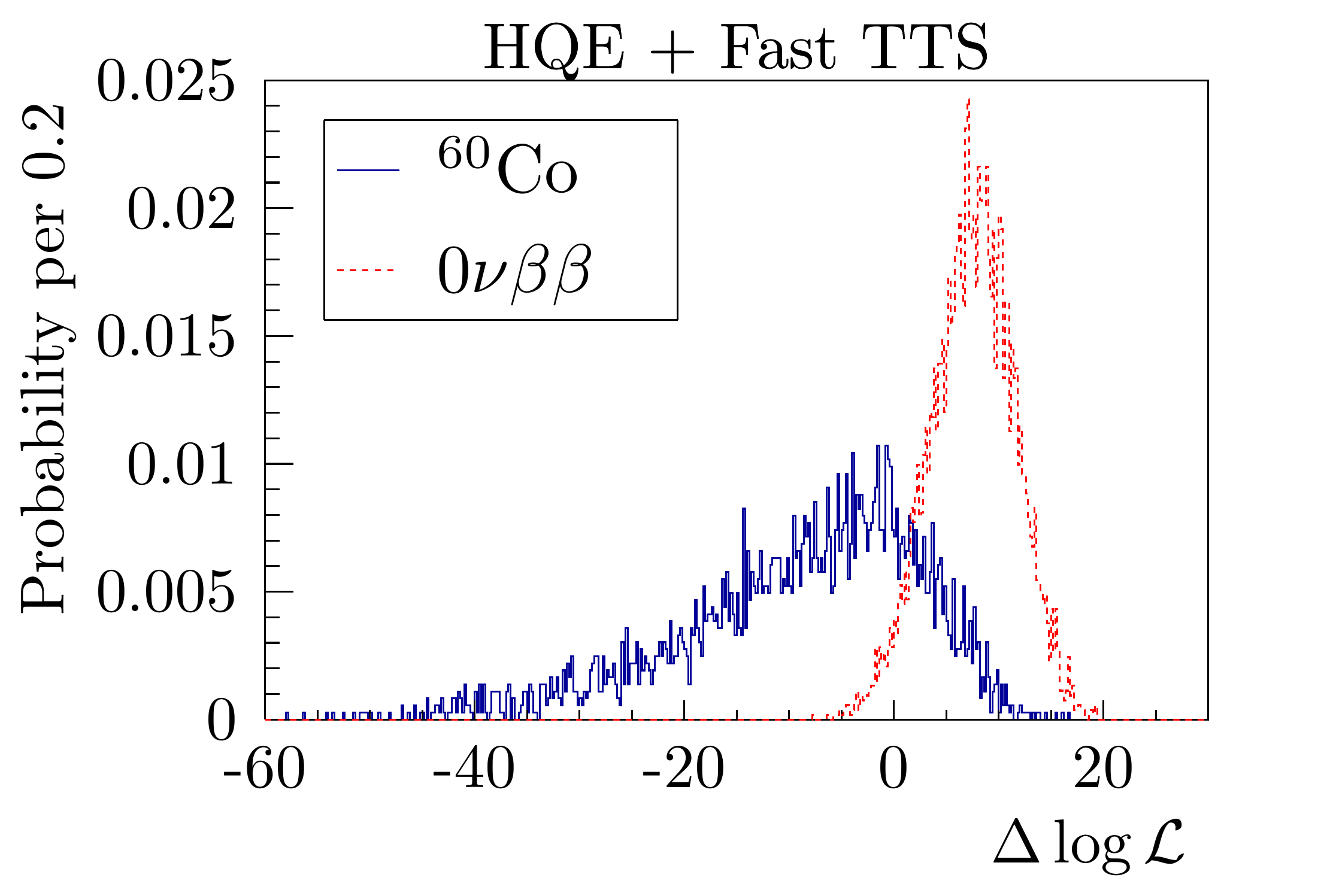} \\
			\large{(a)} & \large{(b)}\\
			\includegraphics[width=0.5\textwidth]{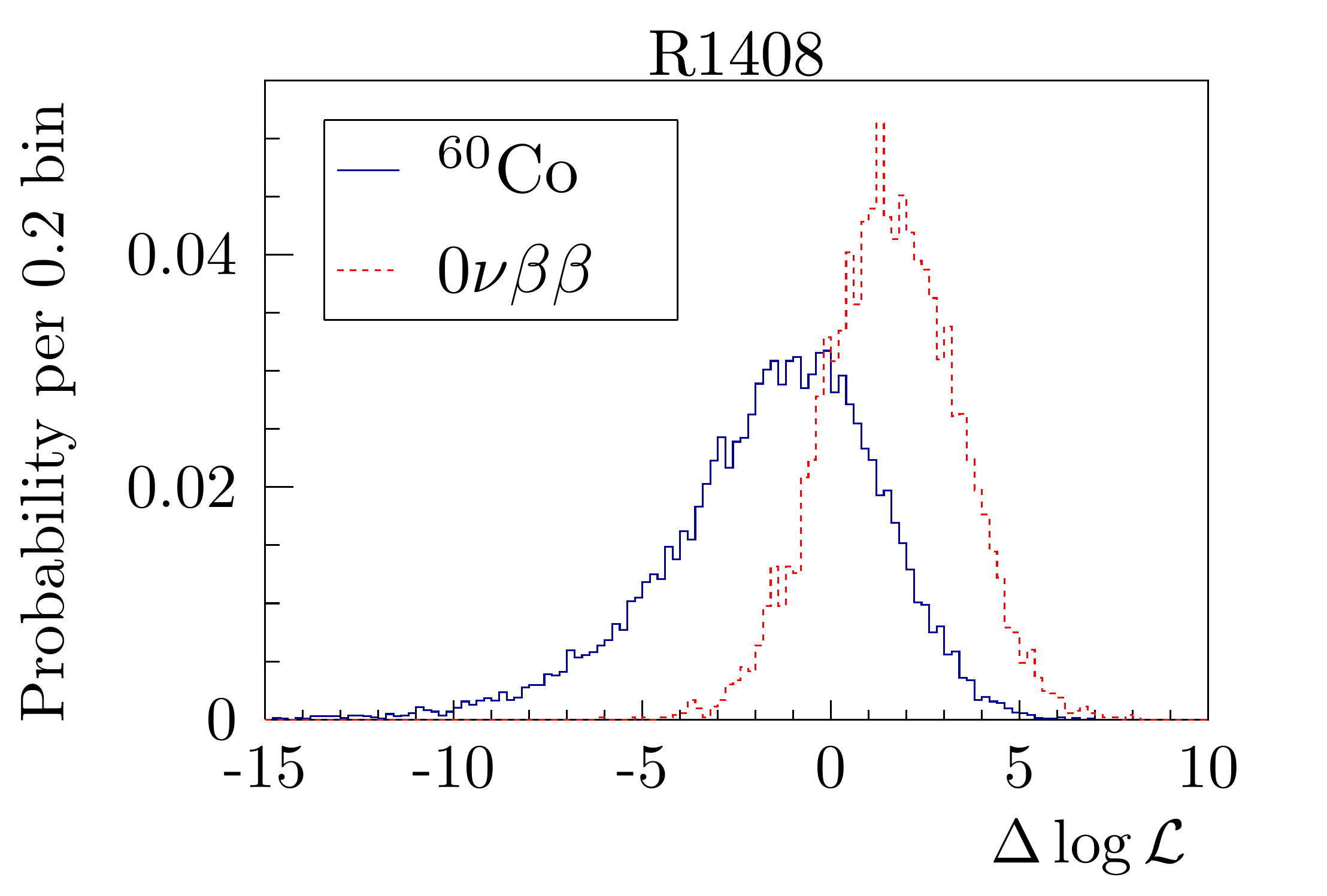} & 
			\includegraphics[width=0.5\textwidth]{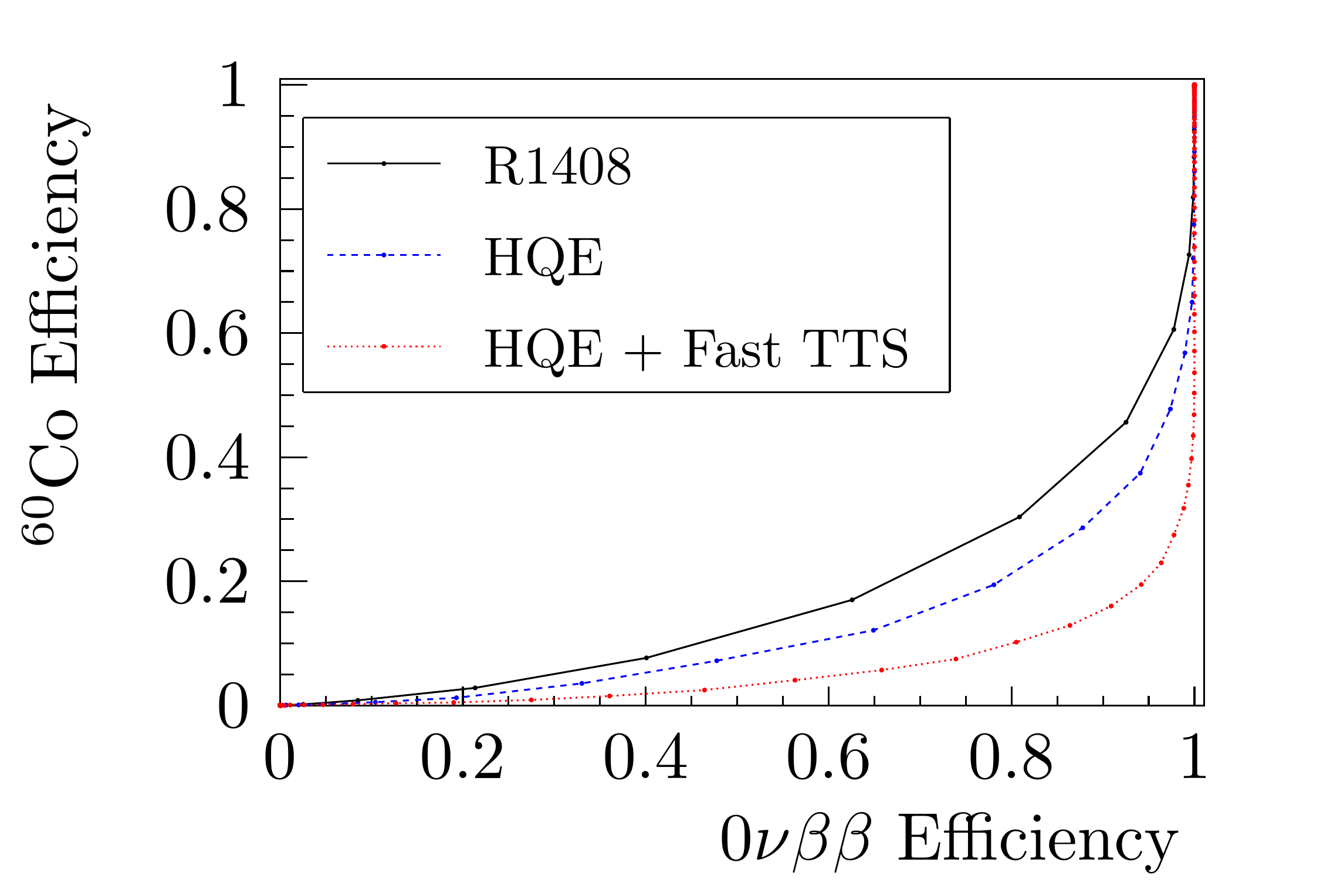} \\
			\large{(c)} & \large{(d)}\\
		\end{tabular}
	}
	\caption{Effect of PMT timing and quantum efficiency on $0\nu\beta\beta$ vs. $^{60}$Co discrimination. (a) using high quantum efficiency PMTs. (b) using high quantum efficiency PMTs with 1ns FWHM TTS. (c) using the Hamamatsu R1408 (d) a comparison of background and signal efficiencies obtainable with a cut on the distributions in (a), (b), (c).}
	\label{fig:upgraded_pmts}
\end{figure}

Figure~\ref{fig:upgraded_pmts} shows the $\Delta\log\mathcal{L}$ distributions with $ S = 0\nu\beta\beta$ and $B = {}^{60}$Co using HQE PMTs and HQE + FastTTS PMTs. It also shows the signal and background efficiencies that could be achieved by cutting on those distributions. As expected, both increased efficiency and reduced TTS improve the separation of the distributions, but the effect is especially marked for the faster PMTs, where event-by-event discrimination becomes efficient enough that one could reject 75\% of $^{60}$Co events with a negligible sacrifice of $0\nu\beta\beta$.

\section{{\it In-situ} Calibration of the technique}

Given the small expected difference between the single-site and multi-site time residual spectra, care must be taken to accurately calibrate both \textit{in-situ}. One way of achieving this is through the deployment tagged calibration sources. An alternative approach, described here, makes use of backgrounds that naturally occur in the detector. 

\subsection{Natural internal sources}

Despite their extreme radio-purity, liquid scintillator detectors contain many sources of radioactivity \cite{Andringa:2015tza}. Almost all of these decays involve the emission of one or more $\gamma$ particles, so there are many candidates for calibrating the multi-site response. The most useful of these can be independently tagged using the method of delayed coincidences.

The first examples are the coincidences produced by uranium/thorium chain contaminants in which a $^{212/214}$Bi $\beta\gamma$ decay is followed by a $^{212/214}$Po $\alpha$ decay after 0.3/164$\mu$s. These events can be tagged with high efficiency using the characteristic delay between the two \cite{Andringa:2015tza}. Usefully, the $^{212/214}$Bi decays have broad energy spectra that overlap significantly with the $0\nu\beta\beta$ region of interest for $^{130}$Te and $^{136}$Xe. A second calibration source comes from 2.2MeV $\gamma$ emitted after neutron capture on hydrogen. The biggest source of free neutrons in liquid scintillator is $(\alpha, n)$ events, where $\alpha$ emitted in radioactive decay are absorbed on carbon nuclei, which then decay via neutron emission. $(\alpha, n)$ events can be tagged using the delayed coincidence between the initial $\alpha$ decay and the neutron capture signal, which is delayed by the neutron thermalisation time ($\approx220\mu$s) \cite{Andringa:2015tza}. Finally, one can tag the delay of 3.1 min between parent $^{212}$Bi $\alpha$ decays and daughter $^{208}$Tl $\beta\gamma$ decays. $^{208}$Tl decays dominate the background budget in the centre of the detector with energies 2.5 - 3MeV, so a pure sample should be readily obtainable despite the longer delay time.

\begin{figure}
  	\includegraphics[width=\textwidth]{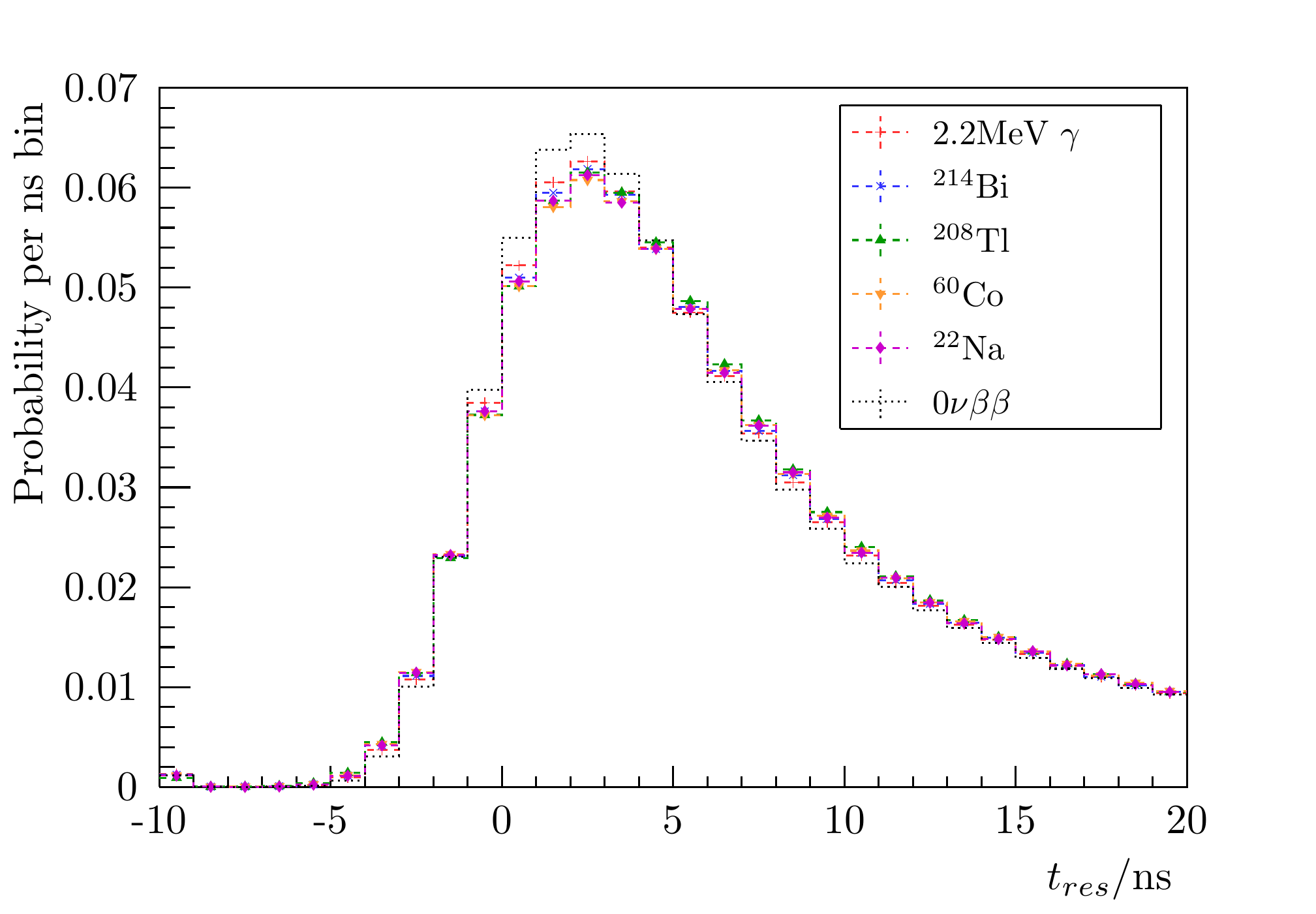}
  	\caption{Time residual spectra for single-site $0\nu\beta\beta$ and multi-site $0\nu\beta\beta$ backgrounds ($^{60}$Co, $^{22}$Na) compared with internal backgrounds that can be tagged via delayed coincidences ($^{208}$Tl, $^{214}$Bi, and 2.2MeV $\gamma$). $r_{fit} < 3.5\textrm{m}$, $3.5\textrm{MeV}  < E_{fit} <  4\textrm{MeV}$ ($^{208}$Tl), $2\textrm{MeV} < E_{fit} <  2.2\textrm{MeV}$ (2.2MeV $\gamma$), $2.31\textrm{MeV}  < E_{fit} <  2.68\textrm{MeV}$ (others).} 
  	\label{fig:ms_tres_calibration}
\end{figure}

Figure~\ref{fig:ms_tres_calibration} shows the time residual spectra for these tagged calibration sources alongside two multi-site backgrounds: $^{60}$Co (a $\gamma$ dominated cosmogenic isotope decay) and $^{22}$Na (a $\beta^+\gamma$ cosmogenic isotope decay). It is clear that the multi-site spectra are much more similar to one another than to single-site $0\nu\beta\beta$, again showing the generality of the technique. If the three tagged samples with different energies can be faithfully reproduced in Monte Carlo, one could reasonably expect to extrapolate to the particles produced by $^{60}$Co, $^{22}$Na etc.

Calibrating the single-site response is more difficult because pure $\beta$ decays are relatively rare at MeV energies and there no delayed coincidences to employ. However, for $0\nu\beta\beta$ experiments using deployed $^{130}$Te or $^{136}$Xe, there will be a strong source of $2\nu\beta\beta$ decays that can be isolated with an energy cut. Near the centre of the SNO+ and KamLAND-ZEN detectors, the $2\nu\beta\beta$ background will dominate over all other backgrounds combined in an energy window just below the $0\nu\beta\beta$ region of interest. 
Simply selecting events in this region would yield a highly enriched sample of $2\nu\beta\beta$ events that could be used to measure the single-site time residual spectrum. An additional source of single-site events comes from solar neutrino elastic scattering events, which dominate the energy spectrum at energies above energies of 3 - 5MeV. Single electron and $0\nu\beta\beta$ events produce scintillation pulses which are  indistinguishable in a SNO+-like detector.

For a calibration of the $\beta^+$ response, one could use muon-following $^{11}$C decays. These events can be tagged using the three-fold-coincidence technique well exploited by Borexino (e.g. \cite{Balata:2006ue}).

\subsection{Further Cross-checks}

Another powerful probe of the multi-site response comes from the distance between the reconstructed position of Bi and Po events in tagged coincidences:

\begin{equation}
 	\Delta R = \| \vec{x}_{fit}^{\textrm{Bi}} - \vec{x}_{fit}^{\textrm{Po}}\|
 	\label{eq:delta_r_def}
\end{equation}
$\Delta R$ is useful because it is related to the track lengths of high energy $\gamma$ emitted in the $^{214}$Bi decay. After a Bi decay, the daughter Po will diffuse a negligible distance during its sub-second lifetime, so the $\beta$ and $\alpha$ decays occur in the same place. Any separation between the two events is therefore dominated by reconstruction $\mathcal{O}$(10cm) or $\gamma$ free streaming (up to 2m). Events with large $\Delta R$ will tend to be from events where the Bi emitted a high energy $\gamma$ that travelled a large distance before scattering. Thus events with large $\Delta R$ are more multi-site, whereas events with small $\Delta R$ will appear to be more single-site.

\begin{figure}
 	\includegraphics[width=\textwidth]{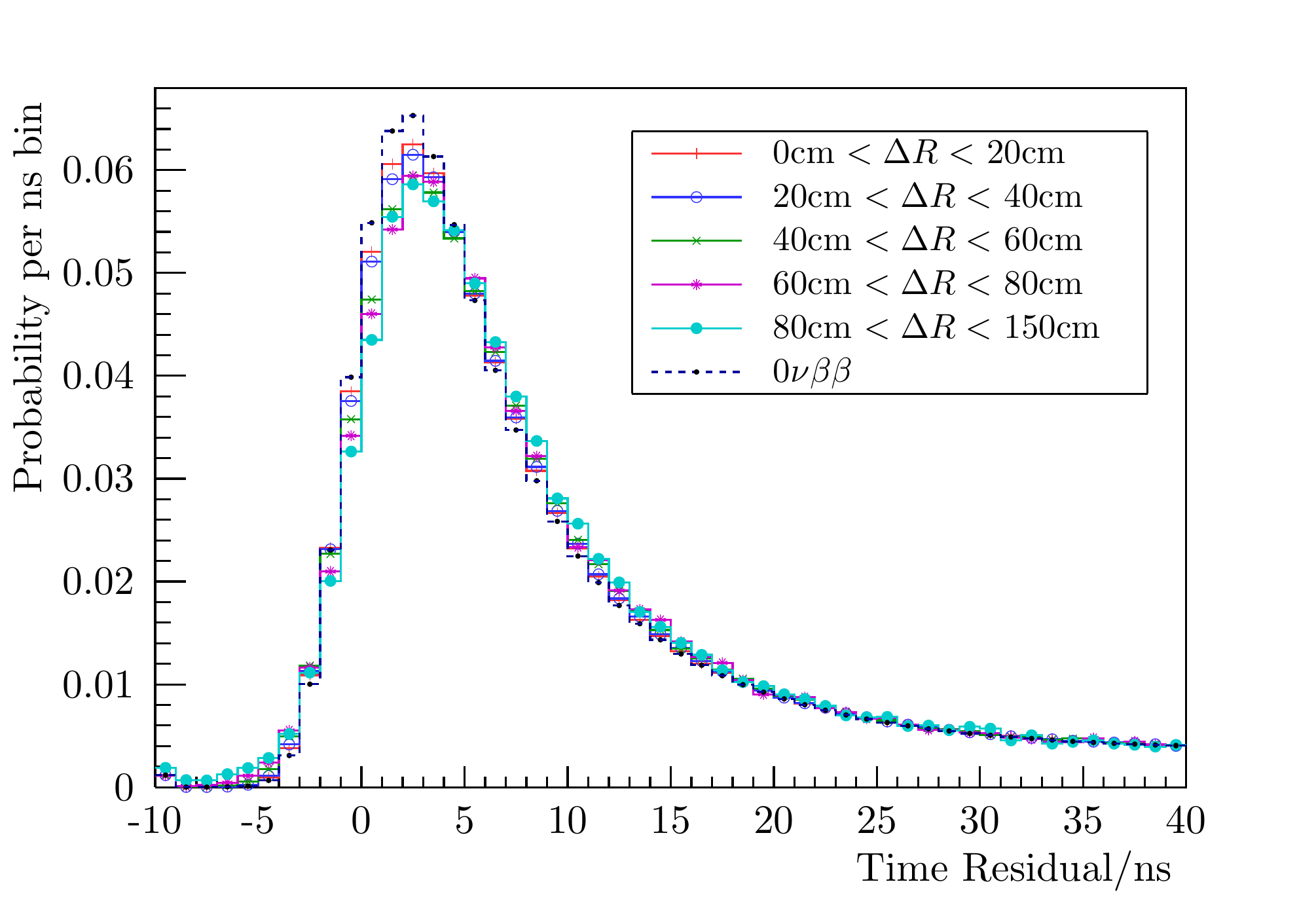}
 	\caption{Time residual distribution for $^{214}$Bi decays binned in $\Delta R$ (the distance between the reconstructed position of the $\beta$ and $\alpha$ decays), alongside $0\nu\beta\beta$.}
 	\label{fig:bipo_pdfs}
 \end{figure}

 \begin{figure}
 	\includegraphics[width=\textwidth]{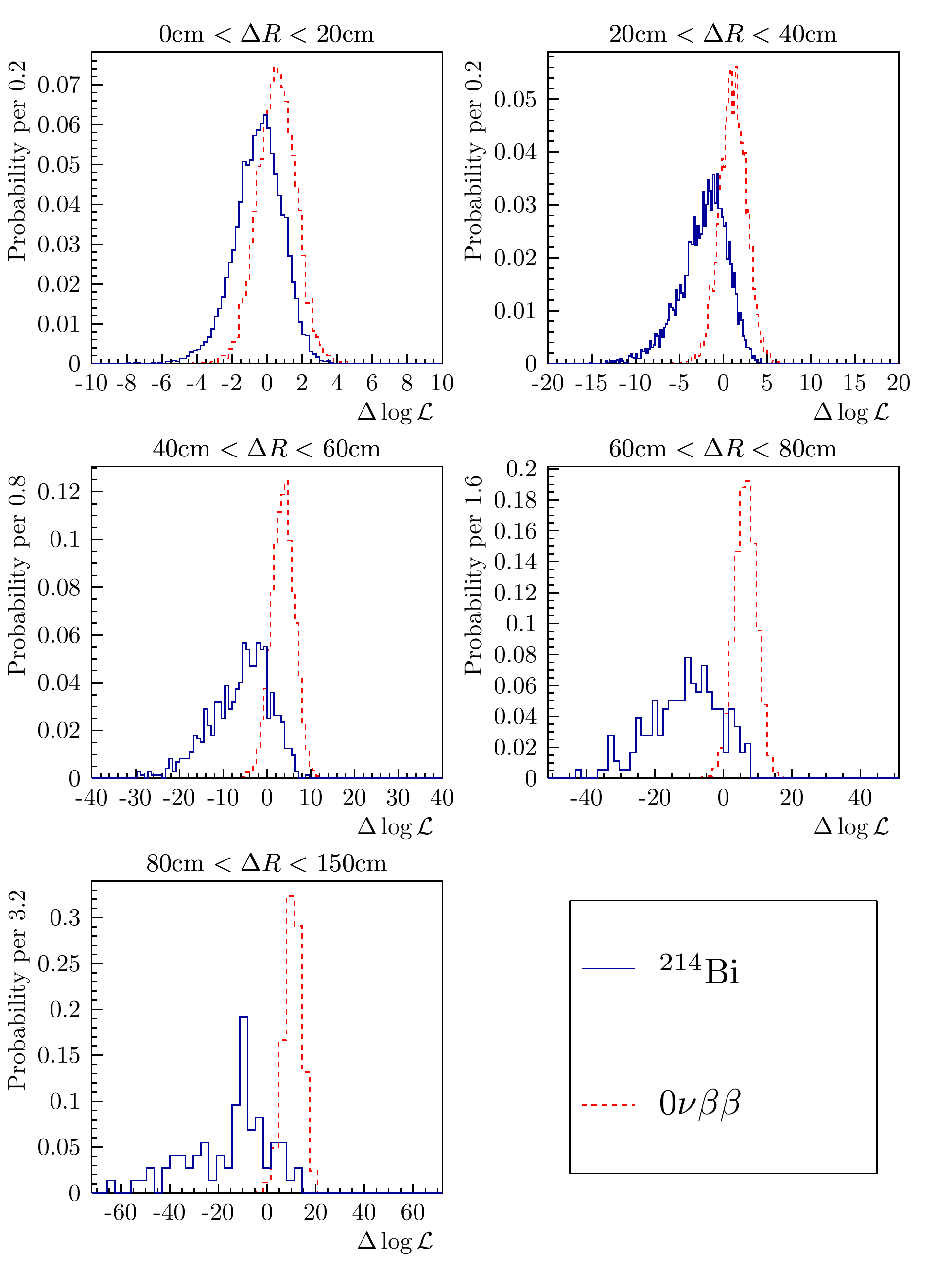}
 	\caption{Likelihood ratios comparing $0\nu\beta\beta$ events with $^{214}$BiPo events binned in $\Delta R$ (the distance between the reconstructed position of the $\beta$ and $\alpha$ decays).}
 	\label{fig:bi214_dr_plot}
\end{figure}

Figures~\ref{fig:bipo_pdfs}~and~\ref{fig:bi214_dr_plot} show the time residual spectra and corresponding $\Delta\log\mathcal{L}$ distributions for $S=0\nu\beta\beta$ and $B={}^{214}$Bi in the 0$\nu\beta\beta$ region of interest. The $0\nu\beta\beta$ distributions in figure~\ref{fig:bi214_dr_plot} have been shown only for comparison with single-site events: their timing response must be calibrated separately. As expected, larger $\Delta R$ events produce broader time residuals and are easier to distinguish from single-site $0\nu\beta\beta$ events. Correctly replicating this behaviour would give a high degree of confidence in the accuracy of the model.


It is also important to calibrate how the single-site and multi-site responses vary with reconstructed energy and reconstructed event radius. This can be achieved using the $^{214/212}$Bi and $2\nu\beta\beta$ samples: both have broad energy spectra that extend up to, or past, the $0\nu\beta\beta$ region of interest and, like $0\nu\beta\beta$, both are distributed throughout the entire detector volume.

\section{\texorpdfstring{0$\nu\beta\beta$}{0nbb} Signal Extraction}

Isotopes like $^{60}$Co, $^{88}$Y and $^{22}$Na can be produced by $\mu$ spallation on Te. These three isotopes are particularly dangerous because they have reconstructed energy spectra that overlap strongly with $0\nu\beta\beta$ and, if present, they are expected to have the same radial distribution as a $0\nu\beta\beta$ signal. Worse still, their half-lives are $\mathcal{O}$(1 yr), which is too long to simply let `cool-down' but short enough to produce a significant decay rate.

Underground purification can almost completely remove these isotopes from the loaded tellurium but, in the event of a positive $0\nu\beta\beta$ signal, one cannot rule out a small contamination of these isotopes without means to distinguish them from $0\nu\beta\beta$. This section examines data from a hypothetical SNO+-like experiment which observes a 3$\sigma$ excess above expected background, asking if, in the presence of a such a signal, the techniques outlined in this paper would allow a true $0\nu\beta\beta$ signal to be discriminated from cosmogenic contamination. It explicitly demonstrates that a fit in energy and radius
alone is insufficient to rule out cosmogenic contamination
and that adding $\Delta\log\mathcal{L}$ as a third fit dimension breaks the degeneracy between signal and background, allowing the the signal to be identified as $0\nu\beta\beta$ with close to the full 3$\sigma$ significance.

To demonstrate this, simulations including many possible backgrounds were produced. The data was then divided into two parts: the first was used to produce PDFs for the fit and the second to produce a representative (`Asimov') data set \cite{Cowan:2010js}, assuming a three year live-time. The scintillator background rates were chosen to match Borexino phase I \cite{Bellini:2013lnn}. The assumed background rates from the loaded tellurium are given in \cite{jackthesis}, they include negligible rates for the cosmogenic isotopes \cite{jackthesis}. A $0\nu\beta\beta$ signal was included, with a size equal to a $3\sigma$ fluctuation above expected background. Several standard analysis cuts were applied to remove mis-reconstructions, BiPo coincidences etc. \cite{jackthesis}.

The $0\nu\beta\beta$ signal was extracted using a Bayesian fit that floats the normalisations of all backgrounds that are expected to contribute more than 1 count/yr in the fit region (1.8MeV $ < E_{fit} < 3$MeV, $r_{fit} < 5.5$m), along with the normalisations of $0\nu\beta\beta$, $^{88}$Y, $^{60}$Co and $^{22}$Na. The joint posterior probability of these normalisations was estimated using a standard extended-likelihood, flat priors for positive normalisations and Markov Chain Monte Carlo. The analysis was performed with and without multi-site discrimination: first, using 2D PDFs in $E_{fit}$ and $r_{corr} = (r_{fit}/6\textrm{m})^3$ and, second, using 3D PDFs with an additional $\Delta\log\mathcal{L}$ dimension, calculated using the $0\nu\beta\beta$ and $^{60}$Co time residual spectra. The PDFs were only extended around the $0\nu\beta\beta$ region of interest. 

\begin{figure}
    \resizebox{\textwidth}{!}{%
        \bgroup
        \def\arraystretch{1.5}
        \begin{tabular}{c c}
            \includegraphics{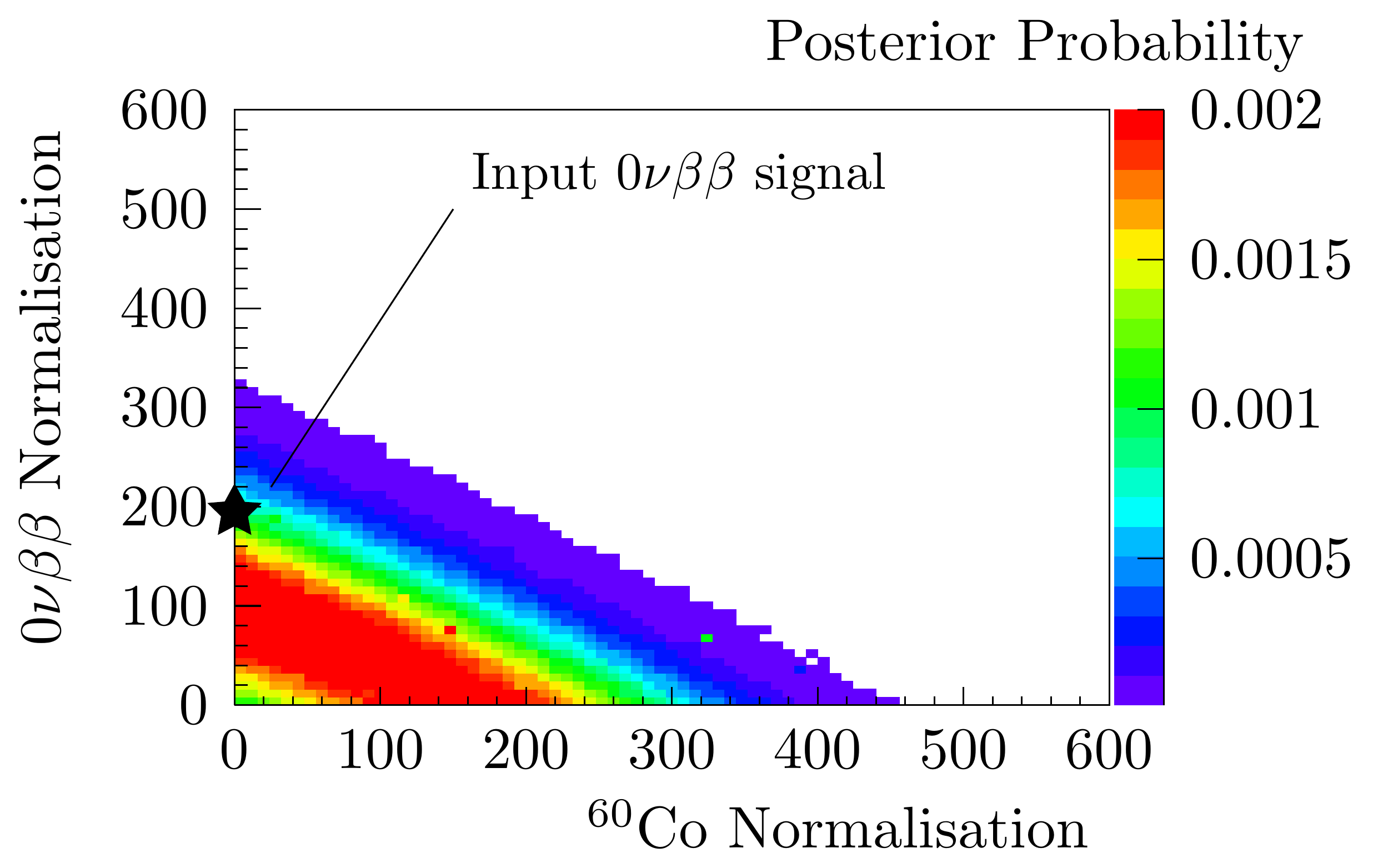} & 
            \includegraphics{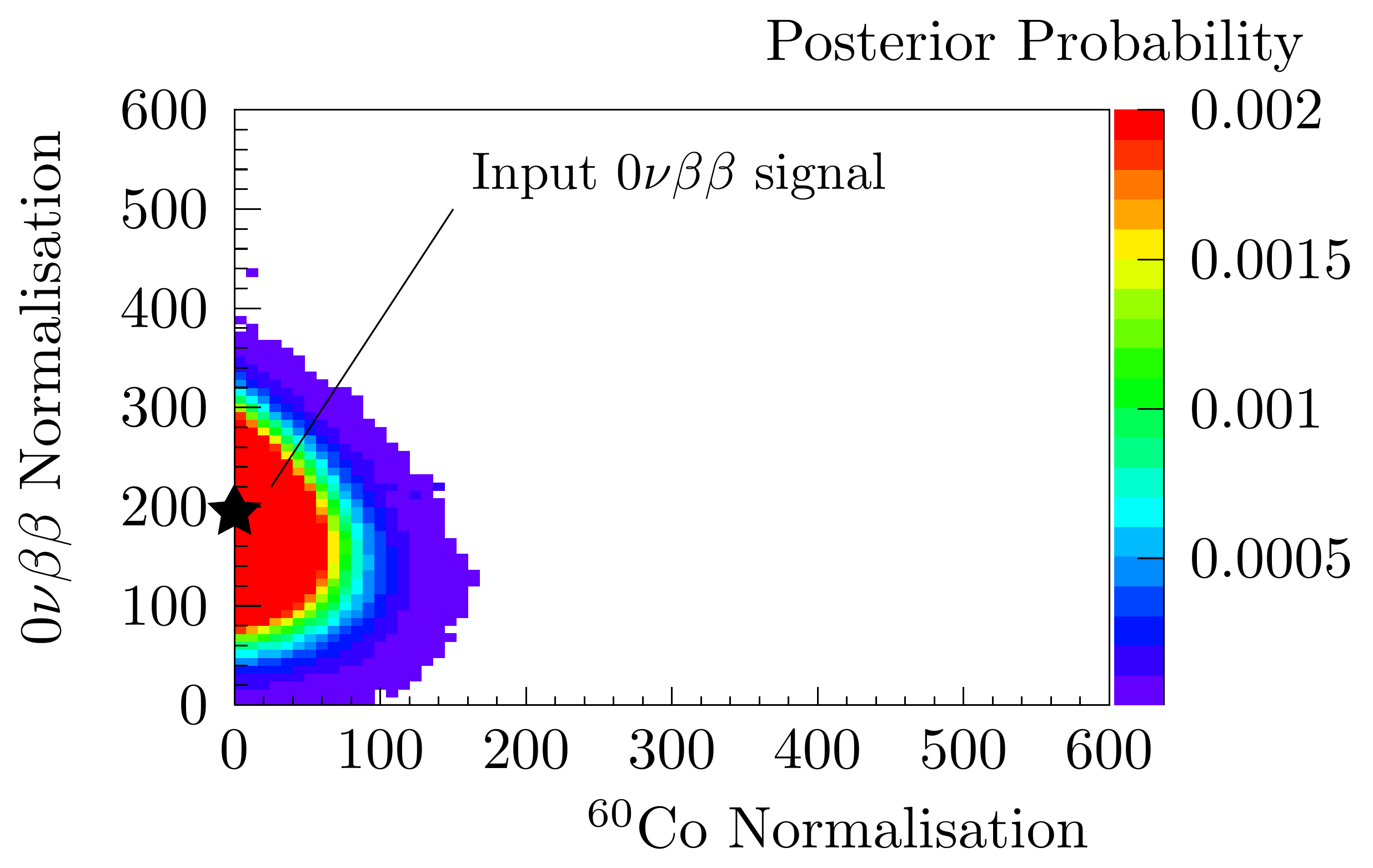} \\
            \includegraphics{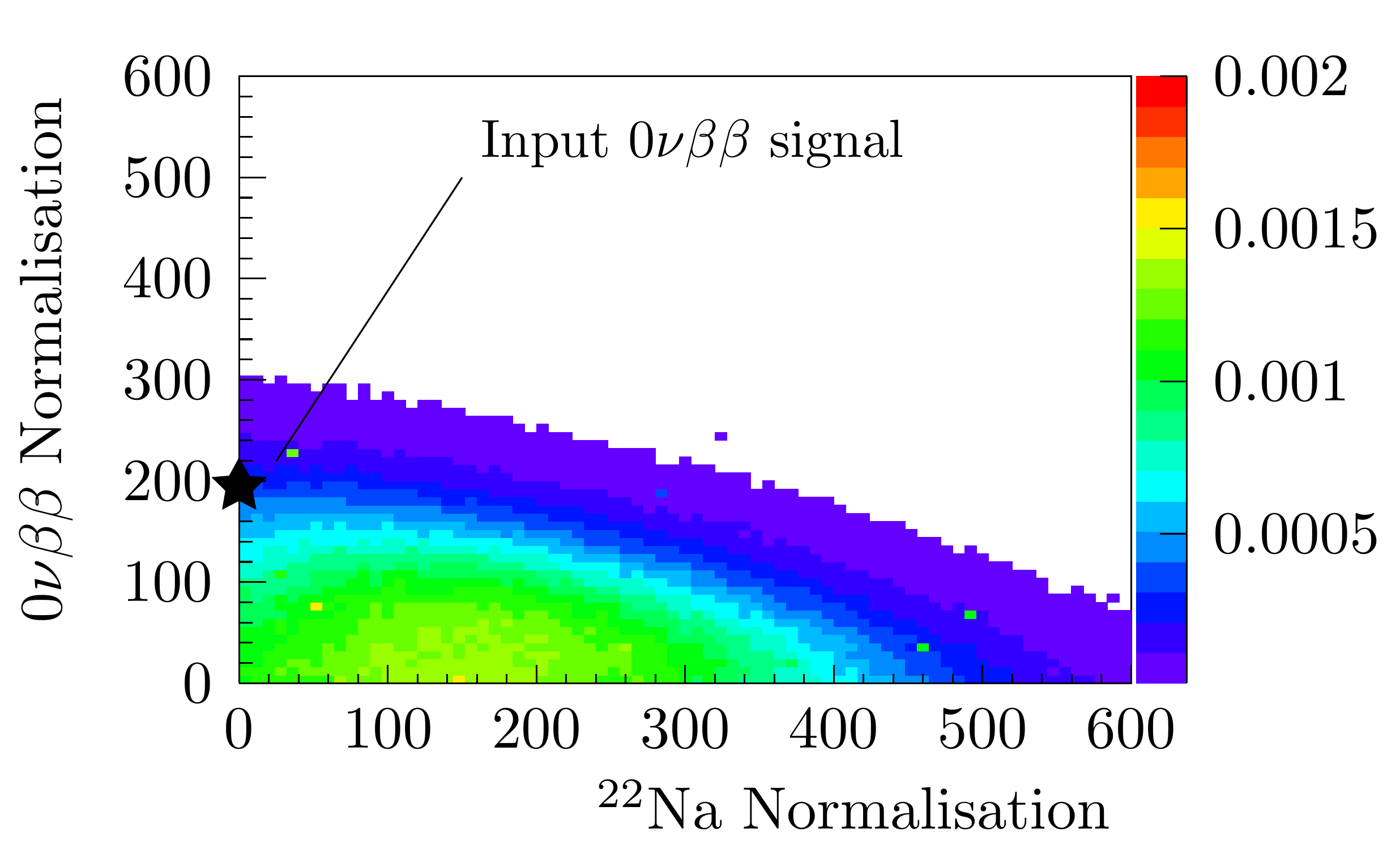} &
            \includegraphics{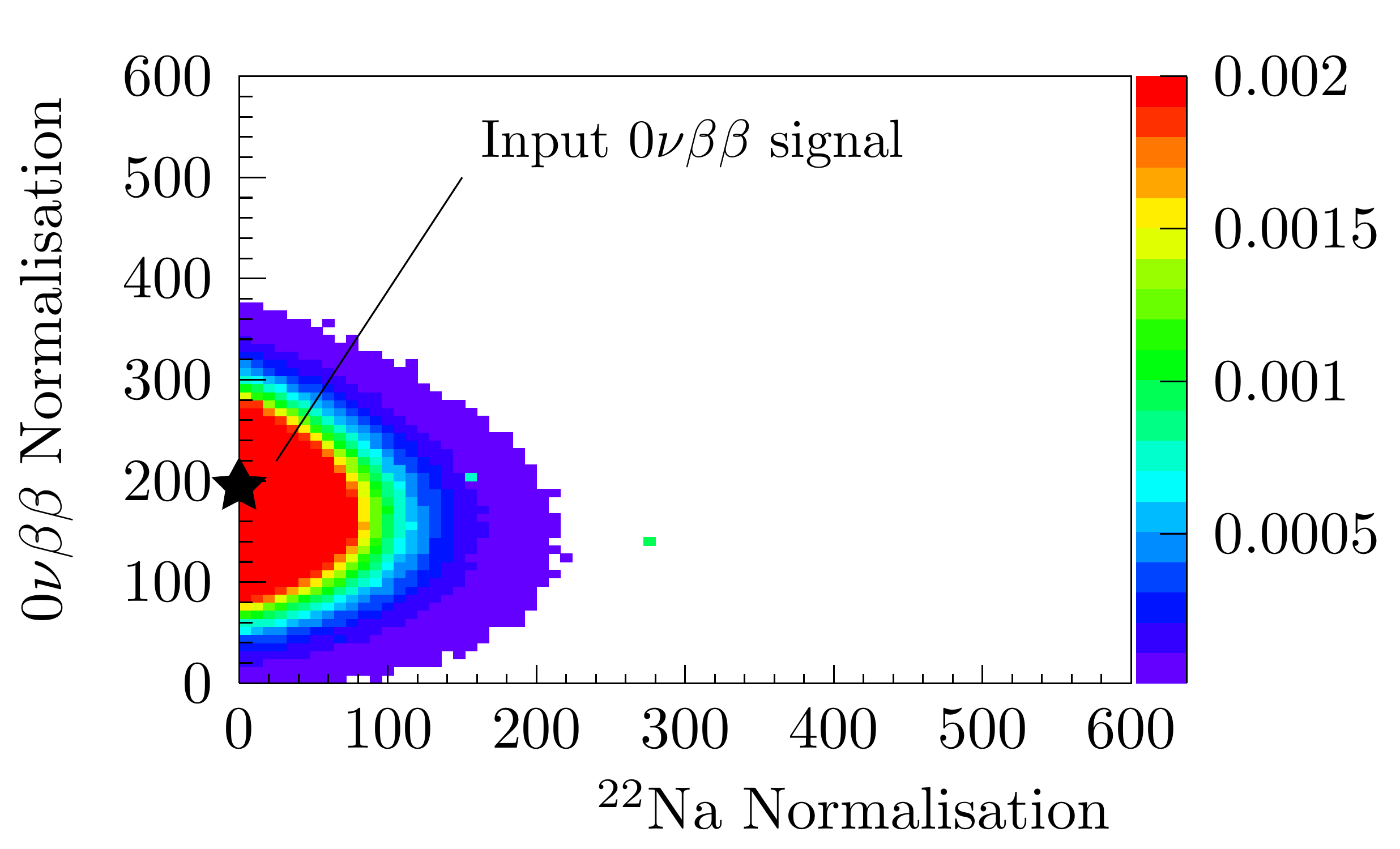} \\
            \includegraphics{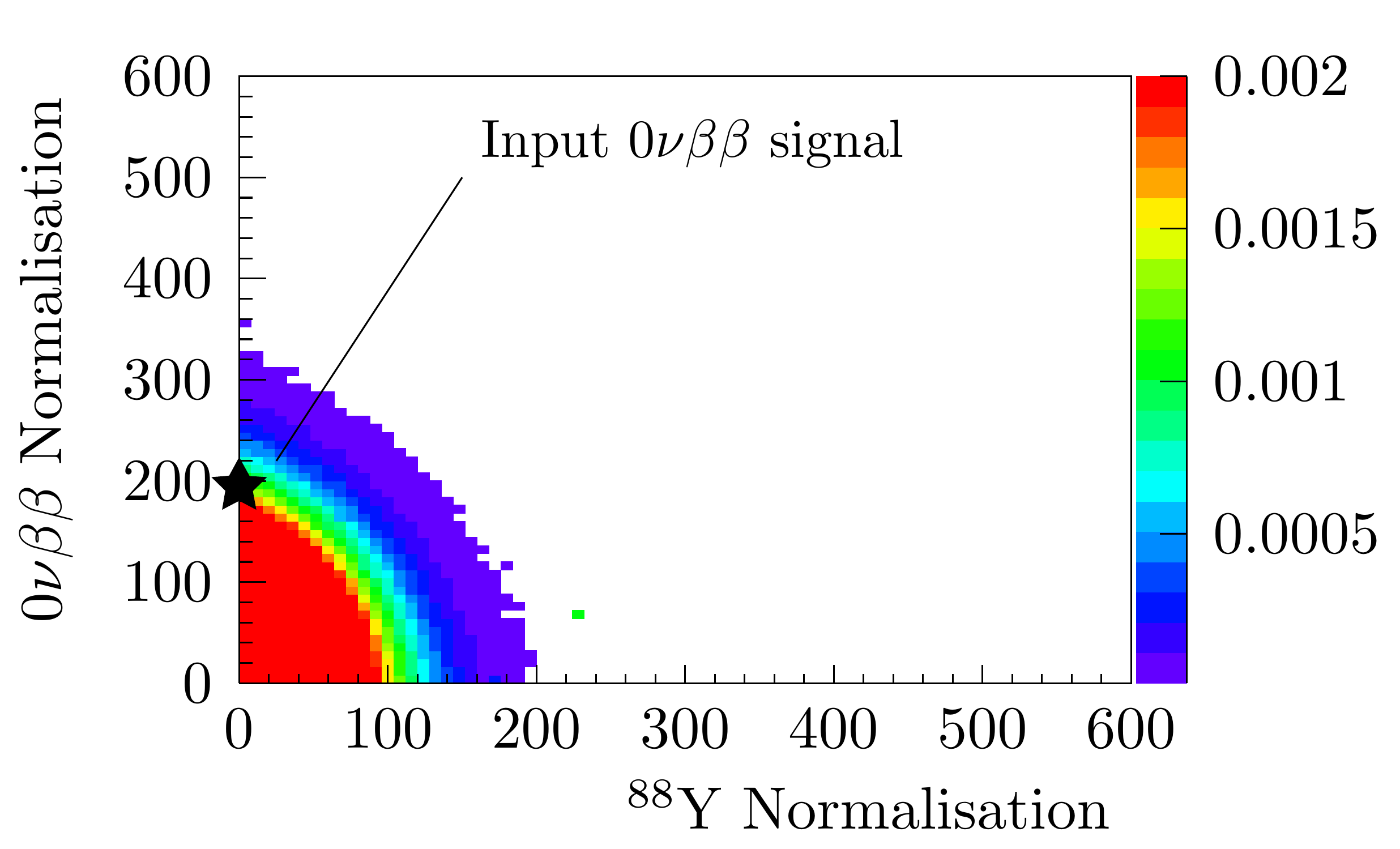} &
            \includegraphics{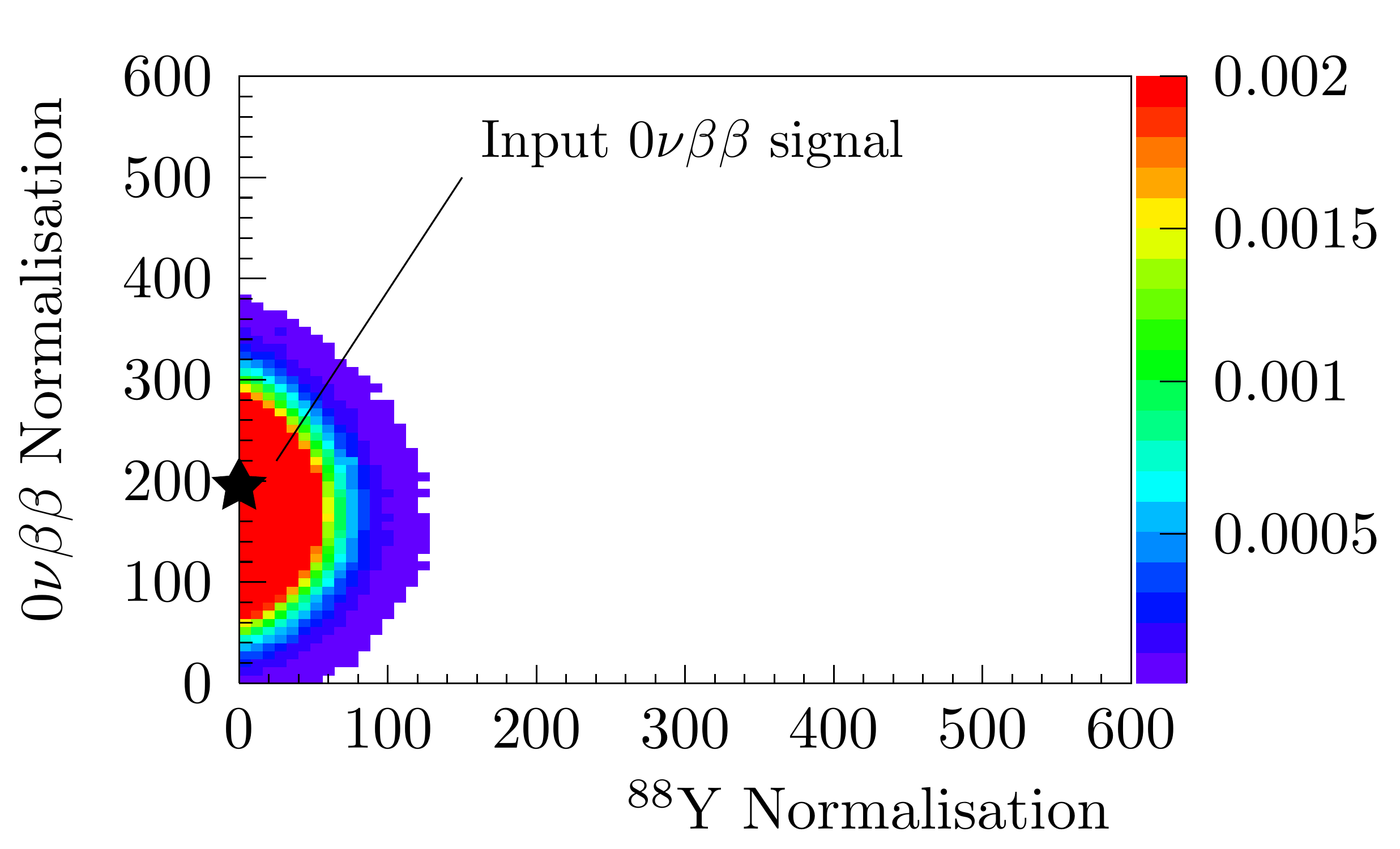} \\
            \includegraphics{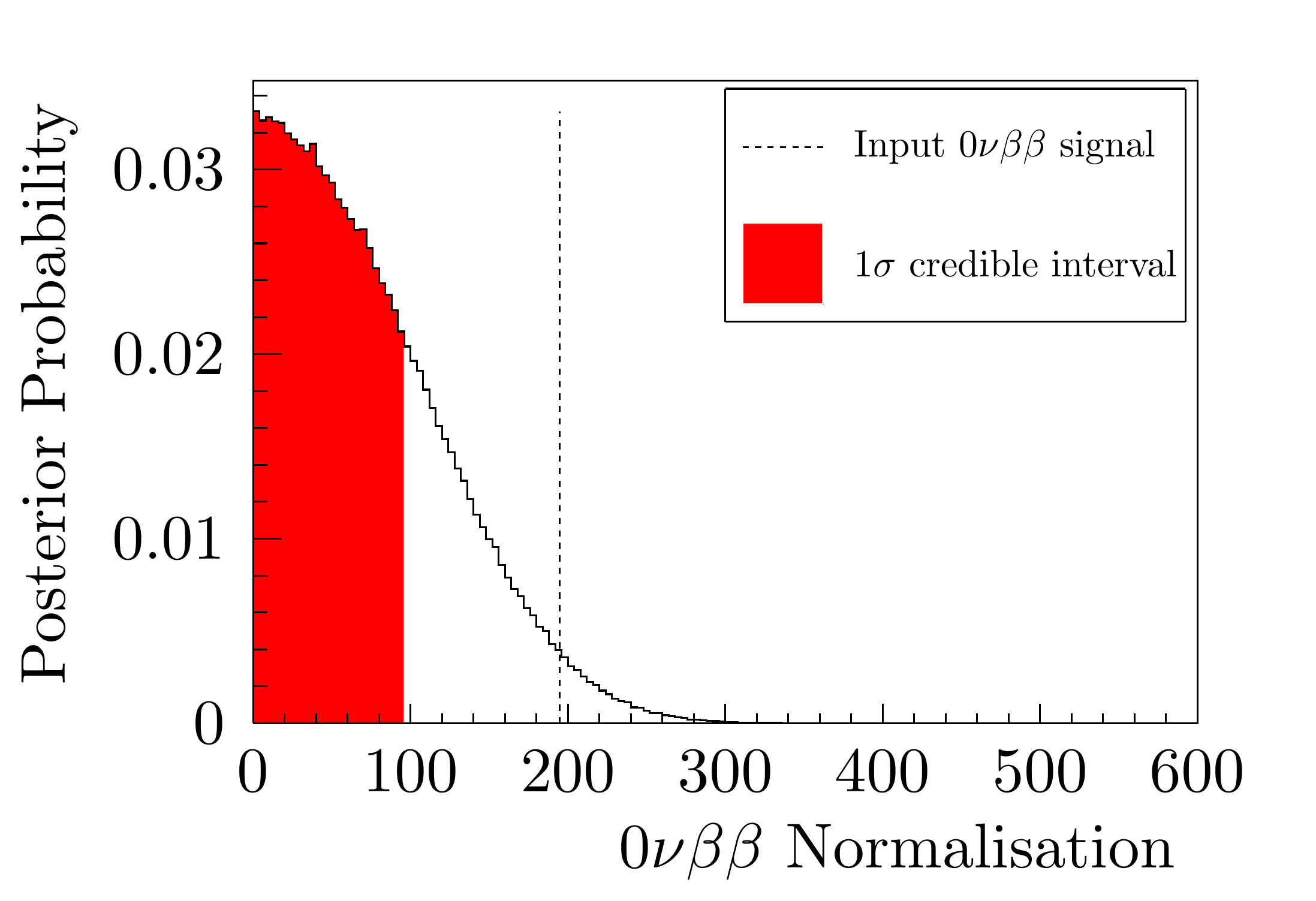} &
            \includegraphics{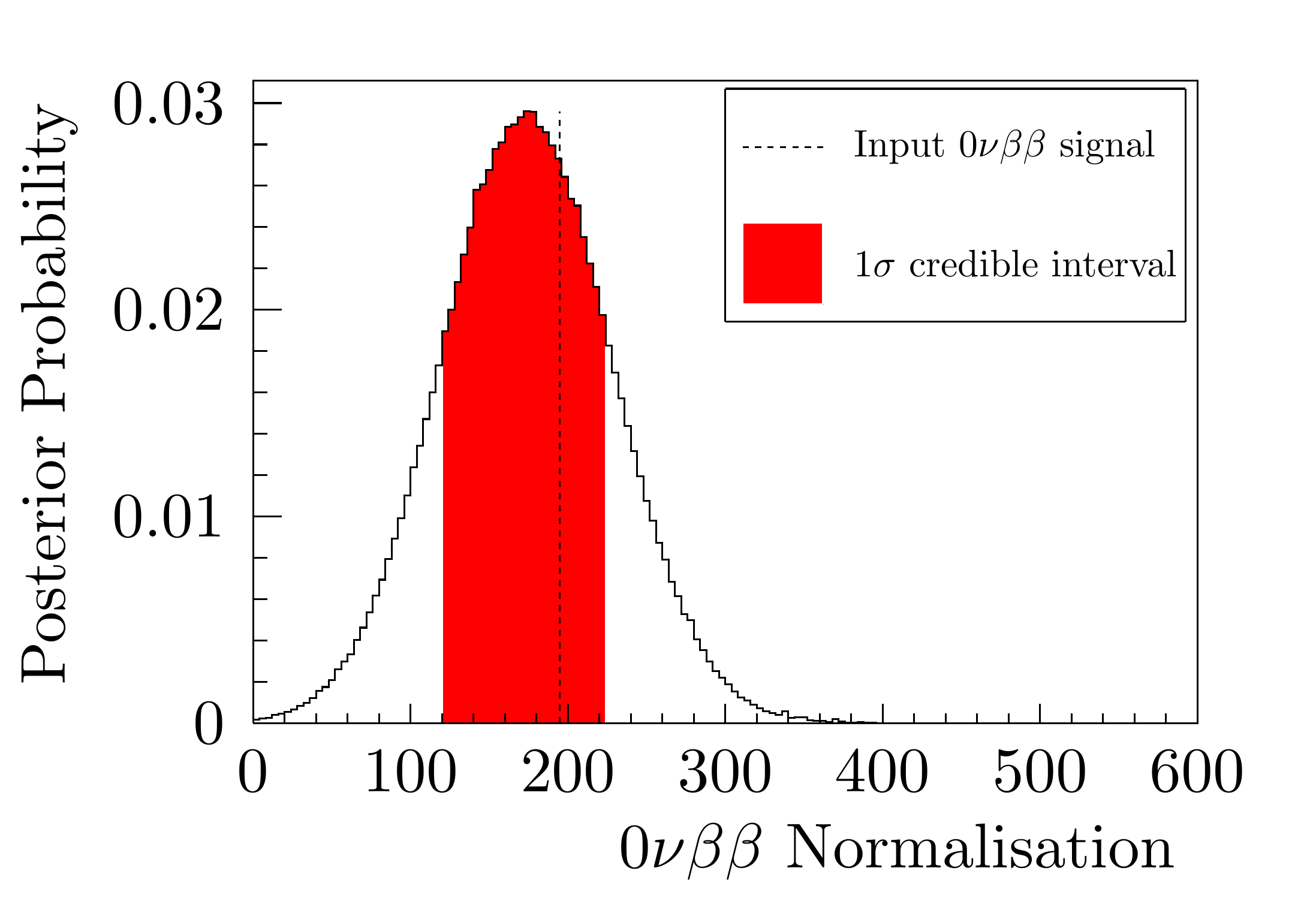} \\
        \end{tabular}
        \egroup
    }
    \caption{(colour on-line) Marginalisations of the posterior probability distribution extracted in a $0\nu\beta\beta$ fit. Left: fit using PDFs in $E_{fit}$ and $r_{corr} = (r_{fit}/6\textrm{m})^3$ alone. Right: using PDFs in $E_{fit}, r_{corr}$ and $\Delta\log\mathcal{L}$, a multi-site discriminator created using the $^{60}$Co and $0\nu\beta\beta$ time residual spectra. Assumes a three year live-time and a 0$\nu\beta\beta$ signal of 262 counts before cuts.}
	\label{fig:degen_breaking}
\end{figure}

Figure~\ref{fig:degen_breaking} shows posteriors extracted from the fits, marginalised to show degeneracies between signal and the cosmogenic backgrounds.
As expected, the 2D fits exhibit strong degeneracies between the signal and the backgrounds, particularly $^{60}$Co. This leads to an underestimation of the $0\nu\beta\beta$ signal, and uncertainties that make the $0\nu\beta\beta$ consistent with 0: without multi-site discrimination, one could not claim a discovery. On the other hand, the 3D fits with multi-site discrimination show no such degeneracy and a positive signal can be claimed with close to the $3\sigma$ significance possible with perfect cosmogenic constraints. The slight underestimation of the signal is due to the finite statistics used to create the Asimov dataset and the PDFs.

This work indicates that the light collection and timing resolution of modern liquid scintillator detectors makes them sensitive to the characteristics of multi-site deposition for events involving $\gamma$ and $\beta^+$, including cascade decays from internal uranium/thorium chain contamination and cosmogenic activity. Furthermore, the multi-site nature of unwanted radioactive backgrounds involving $\gamma$ and $\beta^+$ allows them to be statistically separated from the single-site electron-like signals of interest in $0\nu\beta\beta$ and solar neutrino experiments. A reasonable degree of event-by-event separation is also possible for detectors with larger light collection and/or faster PMTs. The technique is insensitive to the particulars of individual decays and can be calibrated using many \textit{in-situ} backgrounds.  The consequences for $0\nu\beta\beta$ searches are particularly significant, permitting loaded liquid scintillation detectors to become true `discovery' experiments by breaking key degeneracies with potential backgrounds.

\section*{Acknowledgements}

This work was supported by the Science Technology and Facilities Council of the United Kingdom. The authors would like to thank the SNO+ collaboration for use of their \texttt{RAT} simulation, scintillator model and decay generator, as well as many fruitful discussions.

\bibliography{ms}

\begin{thebibliography}{20}
\expandafter\ifx\csname natexlab\endcsname\relax\def\natexlab#1{#1}\fi
\providecommand{\url}[1]{\texttt{#1}}
\providecommand{\href}[2]{#2}
\providecommand{\path}[1]{#1}
\providecommand{\DOIprefix}{doi:}
\providecommand{\ArXivprefix}{arXiv:}
\providecommand{\URLprefix}{URL: }
\providecommand{\Pubmedprefix}{pmid:}
\providecommand{\doi}[1]{\href{http://dx.doi.org/#1}{\path{#1}}}
\providecommand{\Pubmed}[1]{\href{pmid:#1}{\path{#1}}}
\providecommand{\bibinfo}[2]{#2}
\ifx\xfnm\relax \def\xfnm[#1]{\unskip,\space#1}\fi
\bibitem[{Albert et~al.(2014)}]{exo_ms}
\bibinfo{author}{J.~B. Albert}, et~al. (\bibinfo{collaboration}{EXO}),
\newblock \bibinfo{title}{Improved measurement of the
  $2\ensuremath{\nu}\ensuremath{\beta}\ensuremath{\beta}$ half-life of
  ${}^{136}${Xe} with the {EXO}-200 detector},
\newblock \bibinfo{journal}{Phys. Rev. C} \bibinfo{volume}{89}
  (\bibinfo{year}{2014}) \bibinfo{pages}{015502}.
\bibitem[{Agostini et~al.(2013)}]{Agostini2013}
\bibinfo{author}{M.~Agostini}, et~al. (\bibinfo{collaboration}{Gerda}),
\newblock \bibinfo{title}{Pulse shape discrimination for {G}erda phase {I}
  data},
\newblock \bibinfo{journal}{The European Physical Journal C}
  \bibinfo{volume}{73} (\bibinfo{year}{2013}) \bibinfo{pages}{2583}.
\bibitem[{Bellini et~al.(2014)}]{Bellini:2013lnn}
\bibinfo{author}{G.~Bellini}, et~al. (\bibinfo{collaboration}{Borexino}),
\newblock \bibinfo{title}{{Final results of Borexino Phase-I on low energy
  solar neutrino spectroscopy}},
\newblock \bibinfo{journal}{Phys. Rev.} \bibinfo{volume}{D89}
  (\bibinfo{year}{2014}) \bibinfo{pages}{112007}.
\bibitem[{Dunger(2017)}]{Dunger_2017}
\bibinfo{author}{J.~Dunger},
\newblock \bibinfo{title}{Pulse shape analysis techniques in liquid
  scintillator for the identification and suppression of radioactive
  backgrounds to neutrinoless double beta decay},
\newblock \bibinfo{journal}{Journal of Physics: Conference Series}
  \bibinfo{volume}{888} (\bibinfo{year}{2017}) \bibinfo{pages}{012083}.
\bibitem[{Li et~al.(2018)Li, Elagin, Fraker, Grant, and Winslow}]{Li:2018rzw}
\bibinfo{author}{A.~Li}, \bibinfo{author}{A.~Elagin},
  \bibinfo{author}{S.~Fraker}, \bibinfo{author}{C.~Grant},
  \bibinfo{author}{L.~Winslow}, \bibinfo{title}{{Suppression of Cosmic Muon
  Spallation Backgrounds in Liquid Scintillator Detectors Using Convolutional
  Neural Networks}}, \bibinfo{year}{2018}.
  \href{http://arxiv.org/abs/1812.02906}{\tt arXiv:1812.02906}.
\bibitem[{Andringa et~al.(2016)}]{Andringa:2015tza}
\bibinfo{author}{S.~Andringa}, et~al. (\bibinfo{collaboration}{SNO+}),
\newblock \bibinfo{title}{{Current Status and Future Prospects of the SNO+
  Experiment}},
\newblock \bibinfo{journal}{Adv. High Energy Phys.} \bibinfo{volume}{2016}
  (\bibinfo{year}{2016}) \bibinfo{pages}{6194250}.
\bibitem[{Boger et~al.(2000)}]{Boger:1999bb}
\bibinfo{author}{J.~Boger}, et~al. (\bibinfo{collaboration}{SNO}),
\newblock \bibinfo{title}{{The Sudbury neutrino observatory}},
\newblock \bibinfo{journal}{Nucl. Instrum. Meth.} \bibinfo{volume}{A449}
  (\bibinfo{year}{2000}) \bibinfo{pages}{172--207}.
\bibitem[{Amaudruz et~al.(2019)}]{TheDEAP:2017bxf}
\bibinfo{author}{P.~A. Amaudruz}, et~al. (\bibinfo{collaboration}{DEAP}),
\newblock \bibinfo{title}{{In-situ characterization of the Hamamatsu R5912-HQE
  photomultiplier tubes used in the DEAP-3600 experiment}},
\newblock \bibinfo{journal}{Nucl. Instrum. Meth.} \bibinfo{volume}{A922}
  (\bibinfo{year}{2019}) \bibinfo{pages}{373--384}.
\bibitem[{Kaptanoglu(2018)}]{Kaptanoglu:2017jxo}
\bibinfo{author}{T.~Kaptanoglu},
\newblock \bibinfo{title}{{Characterization of the Hamamatsu 8" R5912-MOD
  Photomultiplier Tube}},
\newblock \bibinfo{journal}{Nucl. Instrum. Meth.} \bibinfo{volume}{A889}
  (\bibinfo{year}{2018}) \bibinfo{pages}{69--77}.
\bibitem[{{T. Bolton et al.}(2019)}]{rat}
\bibinfo{author}{{T. Bolton et al.}}, \bibinfo{title}{{RAT} users{{'}} guide,
  \url{https://rat.readthedocs.io/en/latest/}}, \bibinfo{year}{accessed on
  Janurary 1st 2019}.
\bibitem[{{G. A Horton-Smith}(2019)}]{glg4sim}
\bibinfo{author}{{G. A Horton-Smith}}, \bibinfo{title}{{Generic liquid
  scintillator {G}eant4 simulation},
  \url{http://neutrino.phys.ksu.edu/∼GLG4sim/}}, \bibinfo{year}{accessed on
  Janurary 1st 2019}.
\bibitem[{{S. Agostinelli et al.}(2003)}]{geant4}
\bibinfo{author}{{S. Agostinelli et al.}},
\newblock \bibinfo{title}{Geant4 - a simulation toolkit},
\newblock \bibinfo{journal}{Nucl. Instrum. Meth.} \bibinfo{volume}{A506}
  (\bibinfo{year}{2003}) \bibinfo{pages}{250 -- 303}.
\bibitem[{Ponkratenko et~al.(2000)Ponkratenko, Tretyak, and
  Zdesenko}]{Ponkratenko:2000um}
\bibinfo{author}{O.~A. Ponkratenko}, \bibinfo{author}{V.~I. Tretyak},
  \bibinfo{author}{{\relax Yu}.~G. Zdesenko},
\newblock \bibinfo{title}{{The Event generator DECAY4 for simulation of double
  beta processes and decay of radioactive nuclei}},
\newblock \bibinfo{journal}{Phys. Atom. Nucl.} \bibinfo{volume}{63}
  (\bibinfo{year}{2000}) \bibinfo{pages}{1282--1287}. \bibinfo{note}{[Yad.
  Fiz.63,1355(2000)]}.
\bibitem[{Dunger(2018)}]{jackthesis}
\bibinfo{author}{J.~Dunger}, \bibinfo{title}{Topological and Time Based Event
  Classification for Neutrinoless Double Beta Decay in Liquid Scintillator},
  Ph.D. thesis, University of Oxford, \bibinfo{year}{2018}.
\bibitem[{Olive et~al.(2014)}]{Agashe:2014kda}
\bibinfo{author}{K.~A. Olive}, et~al. (\bibinfo{collaboration}{Particle Data
  Group}),
\newblock \bibinfo{title}{{Review of Particle Physics}},
\newblock \bibinfo{journal}{Chin. Phys.} \bibinfo{volume}{C38}
  (\bibinfo{year}{2014}) \bibinfo{pages}{090001}.
\bibitem[{Consolati et~al.(2013)}]{PhysRevC.88.065502}
\bibinfo{author}{G.~Consolati}, et~al.,
\newblock \bibinfo{title}{Characterization of positronium properties in doped
  liquid scintillators},
\newblock \bibinfo{journal}{Phys. Rev. C} \bibinfo{volume}{88}
  (\bibinfo{year}{2013}) \bibinfo{pages}{065502}.
\bibitem[{Coulter(2013)}]{ianthesis}
\bibinfo{author}{I.~Coulter}, \bibinfo{title}{Modelling and reconstruction of
  events in SNO+ related to future searches for lepton and baryon number
  violation}, Ph.D. thesis, University of Oxford, \bibinfo{year}{2013}.
\bibitem[{Lozza and Petzoldt(2015)}]{Lozza:2014haa}
\bibinfo{author}{V.~Lozza}, \bibinfo{author}{J.~Petzoldt},
\newblock \bibinfo{title}{{Cosmogenic activation of a natural tellurium
  target}},
\newblock \bibinfo{journal}{Astropart. Phys.} \bibinfo{volume}{61}
  (\bibinfo{year}{2015}) \bibinfo{pages}{62--71}.
\bibitem[{Back et~al.(2006)}]{Balata:2006ue}
\bibinfo{author}{H.~Back}, et~al. (\bibinfo{collaboration}{Borexino}),
\newblock \bibinfo{title}{{CNO and pep neutrino spectroscopy in Borexino:
  Measurement of the deep underground production of cosmogenic 11C in organic
  liquid scintillator}},
\newblock \bibinfo{journal}{Phys. Rev.} \bibinfo{volume}{C74}
  (\bibinfo{year}{2006}) \bibinfo{pages}{045805}.
\bibitem[{et~al.(2011)}]{Cowan:2010js}
\bibinfo{author}{G.~C. et~al.},
\newblock \bibinfo{title}{{Asymptotic formulae for likelihood-based tests of
  new physics}},
\newblock \bibinfo{journal}{Eur. Phys. J.} \bibinfo{volume}{C71}
  (\bibinfo{year}{2011}) \bibinfo{pages}{1554}. \bibinfo{note}{[Erratum: Eur.
  Phys. J.C73,2501(2013)]}.

\end{thebibliography}

\end{document}